\documentclass[%
 reprint,
showpacs,preprintnumbers,
nofootinbib,
 amsmath,amssymb,
 aps,
 showkeys,
]{revtex4-1}

\usepackage{url}
 \usepackage{multirow}
\usepackage{graphicx}
\usepackage{dcolumn}
\usepackage{bm}

\usepackage{enumitem}

\begin{document}


\title{
Resilience by Structural Entrenchment: \\
Dynamics of Single-Layer and Multiplex Networks \\
Following Sudden Changes to Tie Costs}

\author{Paul E. Smaldino}
\email{paul.smaldino@gmail.com}
\affiliation{
Cognitive and Information Sciences, University of California, Merced
}%

\author{Raissa D'Souza}
\affiliation{
Department of Computer Science and Department of Mechanical and Aerospace Engineering, \\
University of California, Davis
}%

\author{Zeev Maoz}
\affiliation{
Department of Political Science, University of California, Davis
}%


\begin{abstract}
\noindent We examine a model of network formation in single-layer and multiplex networks in which individuals have positive incentives for social ties, closed triangles, and spillover edges. In particular, we investigate the influence of shocks to the network in which the cost of social ties changes after an initial equilibrium. We highlight the emergence of {\em structural entrenchment}: the retention of structural features, such as closed triangles and spillover edges, which are formed under historically different conditions from those currently driving network evolution.
This work has broad implications for understanding path dependence in the structure and dynamics of single-layer and multiplex networks. 

\end{abstract}


\maketitle


\section{Introduction}

The formation and persistence of social ties is dictated by the incentives and opportunities to do so on the part of the individuals involved.  Those individual incentives can shape the emergent structure of the networks that subsequently form. For example, when social ties are preferentially made with already well-connected individuals, the resulting networks exhibit a scale-free structure that is quite different from networks formed at random \cite{barabasi1999scaling}. In the formation of social ties, many types of incentives may operate simultaneously, based on the psychology and economics of social connection, as well as the sociological benefits of participating in a rich network. These incentives include the raw costs or benefits of maintaining social relationships \cite{granovetter_weakties_1973, seeman1996social, holt2010social, cacioppo2009perceived}, the costs or benefits of closing triangles \cite{coleman1988capital, coleman1990social}, and the costs or benefits of having what have been called ``spillover ties"---ties with the same individuals across multiple contexts, which, among other things, can save on transaction costs and provide new social affordances \cite{long_local_1958, hinde1976inter, putnam2000bowling, ashmore2004organizing, smaldinoInPressID}. The kinds of network structures that result from such incentives acting in concert are important to understand, but have not been extensively studied, particularly for cases involving multiplex networks.  

Moreover, incentives for social ties may not be constant over time. 
In some cases, the process of network formation is not path-dependent; ties formed between nodes early on have little impact on later tie-formations. In such cases, the structure of the current network will reflect the present incentives that drive individual behavior. 
On the other hand, consider scenarios in which incentives at one time induce the formation of structural features. These features may be maintained even when the original incentives change, even if they could not have arisen {\em de novo} under the second set of incentives. 
A concrete example of this process involves incentives and costs associated with friendship formation at different stages in life. Young adults may find it beneficial to form friendship triangles (the friend of my friend is my friend), as well as some more isolated friendships. When time constraints increase (due to work, marriage, children, etc.), they will often need to restructure their friendships. In such cases, maintaining existing triangular friendships may be more economical than maintaining the isolated friendships, and thus clustered friend groups will persist. If, however, friendships are initially formed under high time-constraints, there may not be sufficient time to facilitate friendship clusters.
We refer to this type of phenomenon as {\em structural entrenchment}: the persistence of structural features formed under different conditions or incentives than those currently prevailing, which would not  have  formed had the current conditions always existed.

We investigate three questions in the present study:
\begin{enumerate}[nolistsep]
\item How do social networks reorganize following shocks, defined as drastic changes in tie-costs? In particular, \\
\item What is the relationship between different tie-formation incentives and network resilience following shocks? \\
\item To what extent do shocks in one layer of a multiplex network affect the reorganization of another layer of the multiplex?
\end{enumerate}
To do so, we study a dynamic model of social network formation on single-layer and multiplex networks with structural incentives that can vary over time. 
Accordingly, we examine a two-layer multiplex network on which incentives exist for social ties, closed triangles, and spillover ties. We consider changes to incentives in the form of system-wide shocks, such that all individuals in the network experience 
drastic changes to the cost of forming or maintaining social ties. Our model is not meant to reproduce any particular social system, but rather to intuit implications for a broad class of systems. Abstract models, even unrealistic ones, have proven quite valuable in forming intuitions of this sort \cite{Wimsatt:1987aa, epstein08, smaldinoModels}. 

\subsection{Social Ties and Triangles}
Social connections are incentivized in many ways. Social connections provide psychological and health benefits \cite{seeman1996social, holt2010social, cacioppo2009perceived}, and opportunities for cooperation \cite{smaldinoInPressID, apicella2012social, cohen2013tags}, learning \cite{lazer2007networks, derex2015foundations, centola2015social}, and economic activity \cite{granovetter_weakties_1973, jackson2002evolution, schweitzer2009economic}. Consider, for example, social ties in the context of friendship. One friend may provide companionship, information about unfamiliar social conventions, and lodging when traveling far from home. Another friend may help with technical projects and, through conversation, the development of a stronger sense of empathy. Thus we may simply say that social ties can carry benefits. There are, however, limits to how those benefits can accrue \cite{saramaki2014}. One cannot have 10,000 close friends (no matter what some avid social media fans claim), because of the cognitive, temporal, and pragmatic costs associated with maintaining all of those relationships. Furthermore, the benefit to social relationships may have diminishing marginal returns. If you have no friends, making one is of tremendous importance. If you have 40 friends, adding a $41^{\text{st}}$ may carry few benefits unless your new friend brings something quite unique to the table. In our model, we will consider benefits to social ties with diminishing marginal returns. Although many factors influence the value of forming a social tie with one individual rather than another, for simplicity we  assume that, all else equal, the value of a social tie is insensitive to the identities of the individuals involved. 

It can also be of importance that one's friends are friends with each other. If your relationship with one friend weakens, the other can help repair it. If three of you work well together,  new synergistic forms of cooperation can emerge that are impossible with only two. The point is that there are important benefits to closing triangles---e.g., for your friends to be friends with each other---beyond the first-order benefits to social ties \cite{coleman1988capital, coleman1990social}.  In our models, we consider benefits to closed triangles that exist independent of the direct benefits to social ties.

\subsection{Multiplexity and Spillover Ties}
The majority of research on social networks has been on single-layer networks, defined by a set of nodes and a set of ties between them. 
Yet, the multi-relational nature of human interaction has long been a consideration~\cite{long_local_1958,hinde1976inter,palla2005,cai2005}. 
That is, for a given set of nodes (representing individuals), there may exist multiple contexts for each of which a different set of ties describes the structure of social relationships, and in which each set of ties is known as a layer. Recently, a body of work has arisen to study formal properties of multiplex networks, which both extends traditional network theory to multiplex networks and also explores unique properties of networks with more than one layer and interdependencies between or among layers \cite{lubell2013ecology, vijayaraghavan2015spillover, brummitt2015coupled, kivela2014multilayer, boccaletti2014structure, nicosia2013growing, kim2013coevolution, cardillo2013emergence, gomez2015layer, bianconi2013, battiston2016}. 

As an example of a multiplex, consider a set of individuals for whom we can construct a neighborhood network indicating residential contiguity among people. Two people are connected if they live on the same block. Consider also a friendship network in which people are connected if they are friends. Finally, consider an organizational network in which two people are connected if they participate together in formal social settings such as work or volunteer organizations. 
Individual behaviors on any of these networks are not necessarily independent of the other networks. You might become friends with your neighbors or the people you work with, and in doing so create opportunities that don't exist for friends who aren't neighbors or neighbors who aren't friends. Influence between layers of a multiplex network is sometimes known as {\em spillover} \cite{vijayaraghavan2015spillover}. 
In our model, we consider a spillover effect in a two-layer network: nodes get a bonus from forming a tie with a node in one layer if they already have a tie with the same node in the other layer.

\subsection{Changing Incentives}
The costs for forming or maintaining ties may change dramatically over time. 
The relative cost to forming new social ties may be low for childless urban twenty-somethings, but rather high when some of those individuals grow older and acquire demanding jobs, romantic partners, and children. However, although one may lose some friends as one's time becomes more constrained, one rarely loses all of them. Social relationships formed when younger and more carefree may become structurally entrenched by acquiring additional benefits, such as those enjoyed by a tight-knit group of friends who look out for one another's interests, which can outweigh the increased costs of maintaining relationships later in life when demands on one's time have increased \cite{palchykov2012, hruschka2010}. Although changes to the incentives for forming and maintaining social ties often occur gradually, they can also occur rapidly. For example, the birth of a child, the death of a family member, or the loss of a job can very rapidly alter the incentives for forming and maintaining social relationships. 

Similar dynamics are also possible when the nodes of a network are institutional conglomerates, such as tribes, corporations, or nation-states, rather than individual people. For example, trade agreements between corporations may form under supportive economic conditions, such as those enjoyed among EU nations. Relations of this sort may become structurally entrenched, as when multiple businesses share suppliers or distributors and also trade with one another. Dramatic changes to relational incentives, such as a sudden increase in tariffs, may damage some existing relationships and hinder new ones from forming, but leave intact those that are structurally entrenched. 

We model changes to the cost of social ties, leaving constant the benefits of ties, triangles, and spillover ties. We refer to these changes as {\em shocks}, because they are sudden, system-wide changes to the system. We are interested both in shocks that {\em increase} costs---which may reduce the capacity of the network to maintain structure in the form of social ties---are well as in shocks that {\em decrease} costs---which may increase the capacity of the network to maintain such structure. 
We explore conditions under which the network exhibits resilience and maintains structure after a shock.

\section{Model}

Nodes represent individuals (or agents), and ties represent an ongoing social relationship between those individuals. For simplicity, all edges are assumed to be undirected and unweighted. Our model is adapted from a study by Burger and Buskens \cite{burger2009network}, who explored network formation on a single-layer network in response to incentives for ties and closed triangles. 
In their model, nodes in an empty network could bilaterally add ties when such an addition increased the utility of both parties, and drop ties unilaterally if doing so would increase either node's utility. We extend this to a two-layer multiplex in which there can be additional incentives for spillover ties. We then examine network formation and explore the effects of exogenous {\em shocks}, which occur after the network has reached an equilibrium. A shock is operationalized here as a system-wide change in the cost of social ties. Burger and Buskens \cite{burger2009network} restricted their analysis to small six-node networks. Our analysis differs in that we consider networks of arbitrary size. Our dynamics also differ from theirs in that agents in our model are able to consider in their decisions the total utility resulting from rewiring---that is, simultaneously dropping one tie and adding a different tie--- whereas their model required all individual add or drop actions to be utility-increasing. 

\subsection{Utility}

An agent's utility results from three aspects of the social structure of an individuals' local network. First, {\em ties} have intrinsic benefits and costs. Each agent receives a direct benefit for each tie it holds with another agent. However, maintaining ties is also costly due to constraints on time, attention, and transaction costs \cite{burger2009network, simon1990invariants, smaldino2011institutional}. We assume that benefits accrue linearly with the number of ties, while the costs accrue at a faster rate. Our functional form therefore represents diminishing marginal returns to adding additional social ties. Other functional forms that accomplish similar diminishing marginal returns are of course possible. 

Second, closing {\em triangles} may yield additional benefits. We focus on scenarios in which local network closure is an important form of social capital, such as through reducing the costs of information search and facilitating the coordination on social norms \cite{coleman1988capital, coleman1990social}. In other scenarios, closed triangles may be undesirable, as utility is gained through bridging structural holes \cite{burt1992structural}. Such scenarios are also of interest, but for simplicity we do not consider them in the present analysis. 

Third, we consider the benefit of {\em spillover} ties across layers of the multiplex. Specifically, we  consider scenarios in which having a tie with an individual in multiple layers (or contexts) carries an additional benefit. For example, being friends with your neighbor may carry benefits beyond the sum of benefits from having a friend and having a neighbor. We refer to the benefits and costs of ties, triangles, and spillover in aggregate as the {\em structural incentives} of the network. The basic assumption is that nodes act to maximize their marginal utility, that is, they choose ties that maximize the net benefits from their structural incentives. 

\begin{figure}
\includegraphics[width=0.35\textwidth]{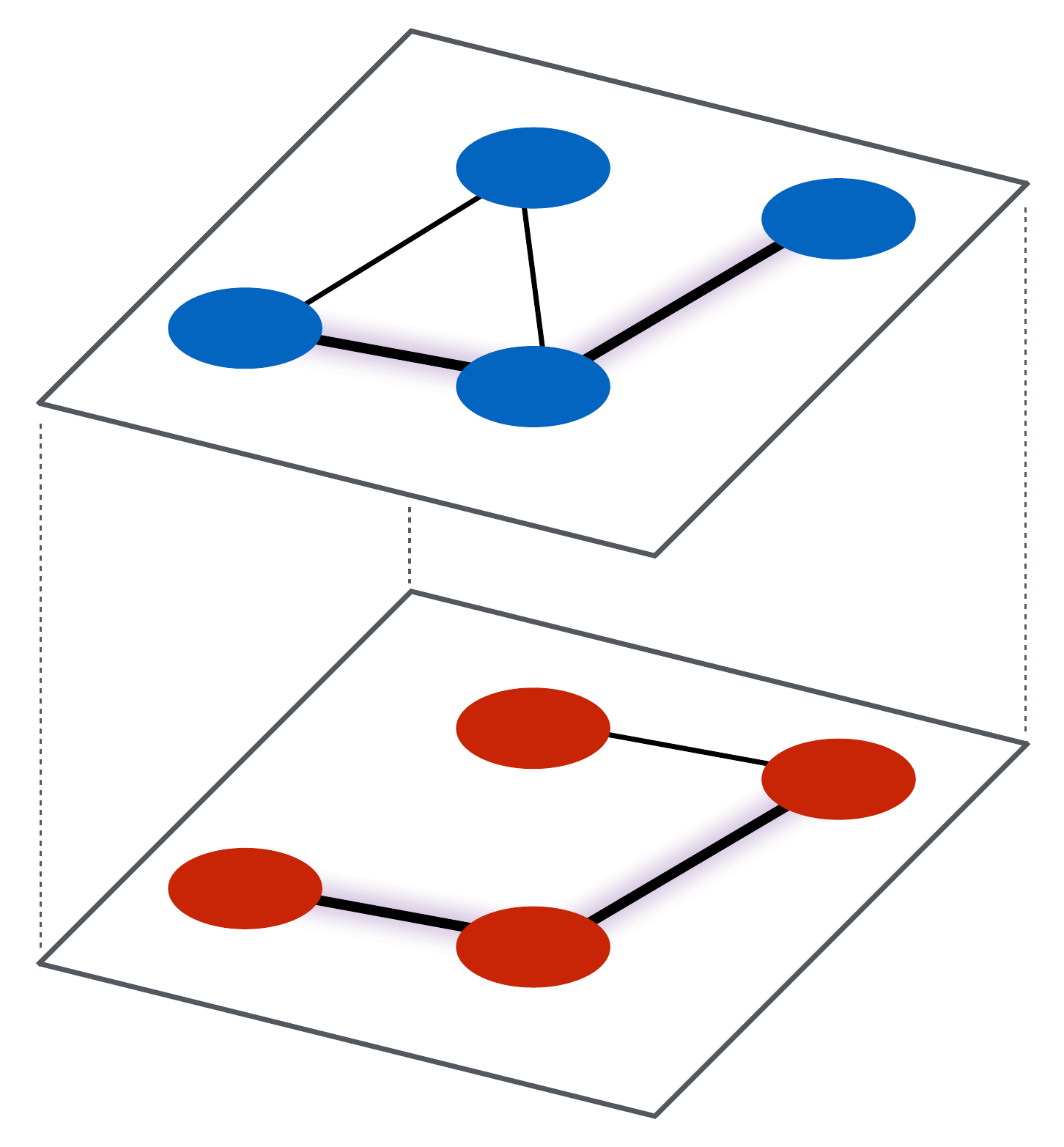}
\caption{\label{fig:fig1} A schematic of the model system, here shown as a four-node multiplex with two layers. The three leftmost nodes are part of a closed triangle in Layer 1 (blue) but not in Layer 2 (red). The three bottommost nodes have spillover ties (ties with the same nodes in both layers), depicted in bold.}
\end{figure}

Our analysis is restricted to a two-layer multiplex (Fig \ref{fig:fig1}). We operationalize utility by extending the functional form introduced in Ref. \cite{burger2009network} to a two-layer multiplex and including spillover benefits. The utility to agent $i$, with $t_{i \ell}$ ties and $z_{i \ell}$ closed triangles in each layer $\ell$ and $v_i$ spillover ties is given by the following function:
\begin{equation}
u_i = \sum_{\ell \in \{1, 2\}} \left( b t_{i \ell} - c t_{i \ell}^2  + d z_{i \ell} \right) + e v_i,
\end{equation}
where $b$ and $c$ are the benefits and costs of maintaining a tie in either layer, $d$ is the benefit to a closed triangle in either layer,  and $e$ is the benefit of spillover ties. 
The benefits to social ties accrue linearly while the costs of social ties accrue quadratically, which operationalizes the idea that the marginal returns to additional social ties will diminish, and eventually become negative, as ties continue to be added (assuming $c < b$). 

For simplicity, we mostly focus on cases in which the structural incentives are the same in each layer, though we do explore one case in which tie costs can vary between layers. Our model is therefore a special case of a more complex model in which each layer has different structural incentives.  Without loss of generality, we set $b = 1$ for all simulations. 

\subsection{Network formation dynamics}

Agents add new ties and drop existing ties in order to increase their utility. Time is discrete and occurs in rounds. 
Each round, each agent has the opportunity to proactively add one new tie and delete one existing tie, though neither action is obligatory. We say ``proactively," because agents may also gain or lose ties through the actions of others. At the beginning of each round, each agent, in random order, samples $p$ other agents in the network. For all our analyses, we use $p = 10$. We keep this number constant across network sizes to reflect constraints on the cognitive and temporal limits to agent observations. 

On its move, an agent $i$ considers all possible ties not already held to each of the $p$ sampled nodes in each layer of the multiplex, and identifies the tie with node $j$ in layer $\ell$ whose addition would provide the largest increase in utility, $\Delta u_{ij \ell}^+$. If multiple ties have equally high value, one is selected at random.  If $\Delta u_{ij \ell}^+ > 0$, agent $i$ proposes the tie. If $\Delta u_{ji \ell}^+ > 0$, that is, if the addition of the tie would also increase $j$'s utility, then the tie is formed, otherwise it is not. Agents can only propose one new tie each round, regardless of whether their proposal is accepted\footnote{Our model assumes that nodes are not aware of the local networks and corresponding utilities of other nodes. If they were, they could selectively offer ties only to those nodes likely to accept them. This informational constraint is likely to apply for some systems and not others.}.

If the straightforward addition of any new tie will not increase the agent's utility, the agent then examines whether it could increase its utility by rewiring, considering only those $p$ nodes sampled. In other words, could the agent increase its utility by dropping a currently held tie with node $h$ and replacing it with a tie with node $j$? Here the agent considers all such pairings, and identifies the pair $(h, j)$ such that dropping its existing edge with $h$ and adding a new tie with $j$ has the largest marginal utility. If that marginal utility is larger than zero, the agent proposes a tie with node $j$. If that tie is acceptable to $j$ (i.e., it increases $j$'s utility), the tie is made, and the agent then drops its edge with node $h$. Otherwise, no action is taken. The newly added tie need not be in the same layer as the dropped tie, corresponding to agents' ability to differentially allocate resources across contexts.  

If no current tie has been dropped, the agent then considers all its current ties, excluding any just added, and identifies the tie for which dropping would lead to the largest marginal utility gain. If that gain is larger than zero, the agent drops the tie. 

This process of network formation continues until a stable network equilibrium has been reached. We operationally define an equilibrium after five complete rounds in which no ties are added or dropped. 

\subsection{Noise}

We focus our analysis on a version of the model in which decisions are deterministic: agents attempt to add only those ties that correspond to the largest gain in utility.  Proposed ties are accepted and existing ties dropped only when they strictly increase an agent's utility. However, we also examine the model's robustness to stochastic noise, governed by the parameter $\nu$. When choosing a new tie to propose, the agent chooses a node to connect with in a utility-maximizing manner (as described above) with probability $1 - \nu$, and with probability $\nu$ the agent selects an (unconnected) node and layer at random. Such proposals are accepted without regard for utility with the same probability. Similarly, an agent drops an existing tie at random with probability $\nu$. Unless otherwise stated, simulations used $\nu = 0$.

\subsection{Shocks}

Once the network reaches a state of equilibrium, a shock occurs. A shock is an exogenous event that simultaneously changes tie costs for all agents. We restrict our analysis to two costs, denoted $c_{low}$ and $c_{high}$. For all simulations, $c_{low} = 0.2$, and, unless otherwise stated, we use $c_{high} = 0.6$. After a shock, new structural changes (i.e., adding new ties or dropping existing ties) may result in a utility increase for some agents. 

Our framework allows for two shock conditions to be compared: low-high (LH) and high-low (HL).  These are contrasted with corresponding control conditions in which no change in cost occurs: low-low (LL), and high-high (HH). The first word (letter) denotes the pre-shock tie cost, and the second word (letter) denotes the post-shock tie cost. 
We examine two variations of shock-related effects. In the first, we examine cases in which changes in costs (i.e., shocks) occur in both layers of the multiplex. In the second, shocks occur only in one layer. This latter variation enables us to study how spillover benefits can cause shocks to propagate across layers. 

Our analysis focuses on the extent to which network structure under the low-cost scenario is maintained after a shock in which tie costs increase---we refer to this extent as the network's {\em resilience}. This is most obviously useful for studying LH shocks, but we find it can also be informative about HL shocks, particularly under mixed effects (in which both triangles and spillover ties are incentivized simultaneously). In the latter case, resilience can be interpreted as the extent to which the network structure under low tie costs can be fully realized when the initial structure evolved under high costs. There are, of course, many ways in which network structure can be characterized. Due to its importance across many areas of network science, its simplicity, and its common interpretation as a measure of network density, we focus on average degree.  
Resilience is operationalized thus:
\begin{equation}
\delta_s = \frac{k_s - k_{HH}}{k_{LL} - k_{HH}},
\end{equation} 
where $k_s$ is the average degree of the network at post-shock equilibrium, and $s \in \{LH, HL\}$ is the shock condition. 
It is also possible to generalize this measure of resilience to {\em any} network-level metric by substituting that metric for average degree, though as noted we have not analyzed other measures of this type.

For most of our multiplex network analyses, the two layers of the multiplex are statistically identical, so for convenience we measure the average degree of Layer 1 only. In cases where the two layers are subjected to different shocks, we compare the average degree of each layer post-shock to Layer 1 of the pre-shock network; in this case each layer will have a unique level of resilience.

\subsection{Single-layer and Multiplex networks}
We were specifically interested in network formation and the effects of spillover in multiplex networks. However, our findings regarding the path dependency of network formation on the timing of shocks has implications on the type of single-layer networks that have more traditionally been studied in network science. Therefore, we also present results on single-layer networks. In these runs, all interactions are restricted to a single layer, and the influence of spillover is undefined (and so $e = 0$). 

Java code for our agent-based model is available at \url{http://www.openabm.org/model/5148/}.

\section{Results}
Here we describe the varieties of network organizations that emerge under the incentives we describe above, under low and high tie costs and after the shocks that take a system at equilibrium from one tie cost to the other. Under a wide range of conditions, much of the network structure facilitated under low costs can be preserved even after the costs significantly increase---structure that could not arise {\em de novo} under high tie costs. As a quantitative metric, we focus on the average degree. For most runs, all incentives were identical for each layer of the multiplex, and so network statistics were effectively identical for each layer. As such, we only present data for Layer 1 of the multiplex unless otherwise indicated. Our simulations cover networks of sizes ranging from 20 to 80 nodes; results are shown in main text are for a 40-node network unless otherwise stated. In the SI Appendix we show that our results are generally similar across these different network sizes (see also Fig. \ref{fig:fig2}).  All data is from 100 simulation runs for each parameter condition unless otherwise stated.

\subsection{Isolated effects}
We first examine single-layer networks (for which no spillover is possible) in which there are additional benefits to closed triangles.  
For our high cost scenario, we purposefully chose an extreme case in which triangles would not emerge due to the prohibitively high costs of maintaining two-stars. However, our main result  is robust, if less stark, for lower tie costs in which triangles do emerge in the high-cost condition (see Figs. \ref{fig:fig2}B,D and SI Appendix). 

Under low costs, many triangles form, and the average degree of the network increases as triangles are incentivized more (Fig. \ref{fig:fig2}A). This is because closed triangles scaffold the creation of addition triangles by providing affordances (e.g., new two-stars), forming a cascade. Such a cascade does not go on indefinitely, as the costs of ties can set a practical limit, especially when each new edge must yield an increase in utility. 
The HL condition closely tracks the LL condition, because both conditions result from dynamics under low tie costs. 

For values of $d$ below a critical threshold, the LH condition tracks the HH condition. That is, there is no resilience by structural entrenchment. This is because the shock in which tie costs increase causes agents to drop ties, and triangles cannot be maintained. Past the critical threshold ($d = 0.8$ in our runs), some amount of resilience occurs. Some edges are dropped following the shock, but the resulting network is denser than networks that began with high tie costs. This first threshold occurs when the benefit of a closed triangle can offset the higher tie cost, so that a node in a closed triangle need not drop any ties. 
Past a second threshold ($d = 1.2$ in our runs), when the benefits to closed triangles are high enough, networks in the LH condition are indistinguishable from networks in the LL condition. See SI Appendix for a derivation of these thresholds and for additional statistics regarding the dynamics of single-layer networks. 


\begin{figure*}
\includegraphics[width=0.8\textwidth]{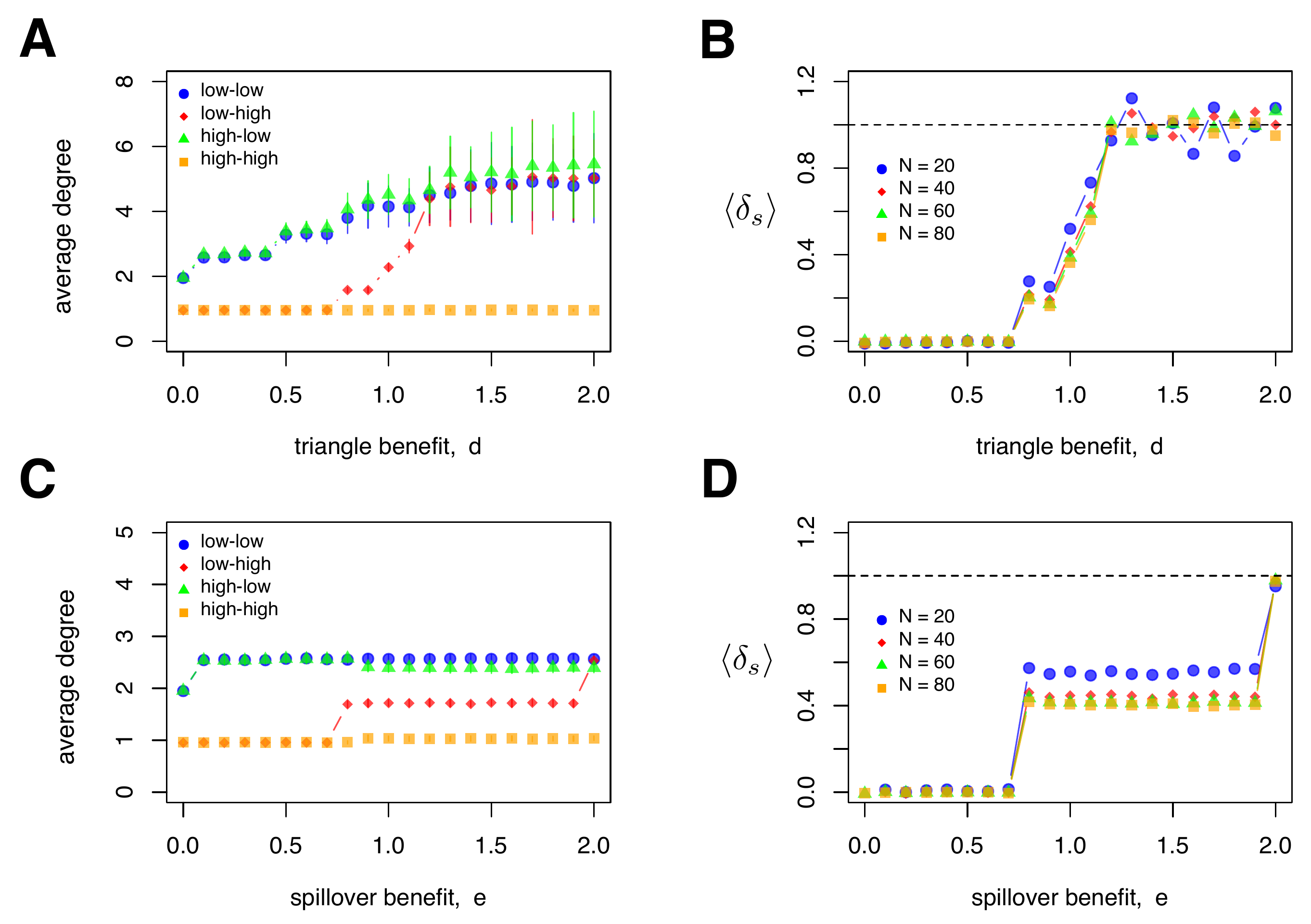}
\caption{\label{fig:fig2} Isolated effects for (A, B) triangle benefits only ($e = 0$) and (C, D) spillover benefits only ($d = 0$). (A, C) Average node degree $\pm$ SD for each of the four shock conditions on a 40-node network. (B, D) Average resilience for LH condition, showing robustness across a range of network sizes.}
\end{figure*}




For all subsequent results, we consider a two-layer multiplex. Like incentives for closed triangles in a single-layer network, incentives for spillover ties can provide a minimal model of structural entrenchment in a multiplex network, as seen in the absence of triangle benefits ($d = 0$). In general, we observe a similar pattern of resilience for spillover as we did for triangles (Fig. \ref{fig:fig2}C). Unlike with triangles, however, the average degree under low tie costs does not continue to increase with the benefit to spillover ties, but rather plateaus. This is because spillover ties do not scaffold the creation of additional spillover ties, as closing triangles does. Nevertheless, our results show that a benefit for spillover ties can facilitate network resilience by structural entrenchment even in the absence of benefits to clustering. 

The network structures that emerge from spillover incentives are quite different from those that emerge from triangle incentives (see SI Appendix). Under low tie costs, incentives for triangles created several tightly clustered but completely discrete communities. Incentives for spillover, on the other hand, tended to create fully connected graphs that exhibit low levels of triadic closure. Additionally, unlike the case of triangle benefits, these networks are not fully resilient until a much higher spillover benefit has been reached as compared with the triangle case ($e = 2$ in our runs; see Fig. \ref{fig:fig2}C). This is because each tie can only confer one unit of spillover benefit, whereas a single tie can be part of many triangles. See SI Appendix for derivation of critical thresholds and for additional analyses on the isolated effects of both triangle and spillover benefits.

\subsection{Mixed effects}
\begin{figure*}
\includegraphics[width=0.99\textwidth]{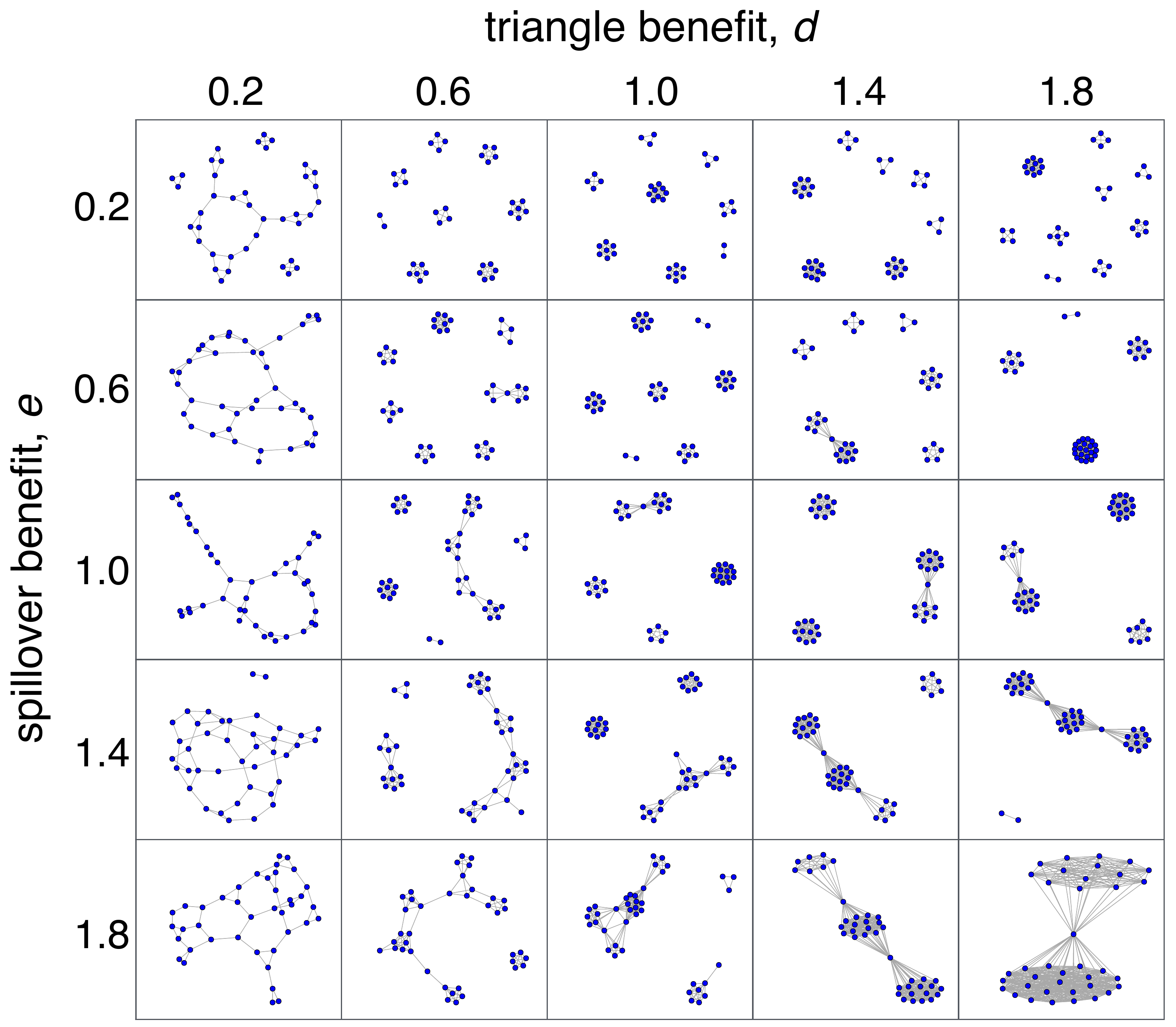}
\caption{\label{fig:fig6}  Representative networks (Layer 1 only) under low costs (LL) that emerge as a result of varying incentives for closed triangles and spillover ties. Unconnected nodes do occasionally occur but are not represented in these plots. These data are based on networks of size $N = 40$; however, similar patterns emerge for networks of larger size ($N$ = 60, 80).}
\end{figure*}

Our focus here is on cases where structural incentives (for triangle closure and spillover) can combine to yield a wide variety of networks.
In Fig. \ref{fig:fig6} we plot representative networks that emerge when both closed triangles and spillover ties are incentivized. 
The combined incentives produce structures that are quite different from what we see when each incentive is considered in isolation (see Figs. S2 and S4 in the SI Appendix). In particular, these combined incentives give rise to large, tightly clustered communities connected via a single brokering node, especially when both incentives are strong. 

Although we study a two-layer multiplex, only one layer is presented in these network diagrams. In all runs presented so far, incentives in each layer are identical, so statistically both layers are identical. However, this does not capture the ways in which the two layers are connected to each other. When spillover benefits are strong relative to triangle benefits, most or all edges will be spillover ties. Otherwise, about half of all ties are spillover edges (see Fig. \ref{fig:fig8}).  

\begin{figure*}
\includegraphics[width=0.99\textwidth]{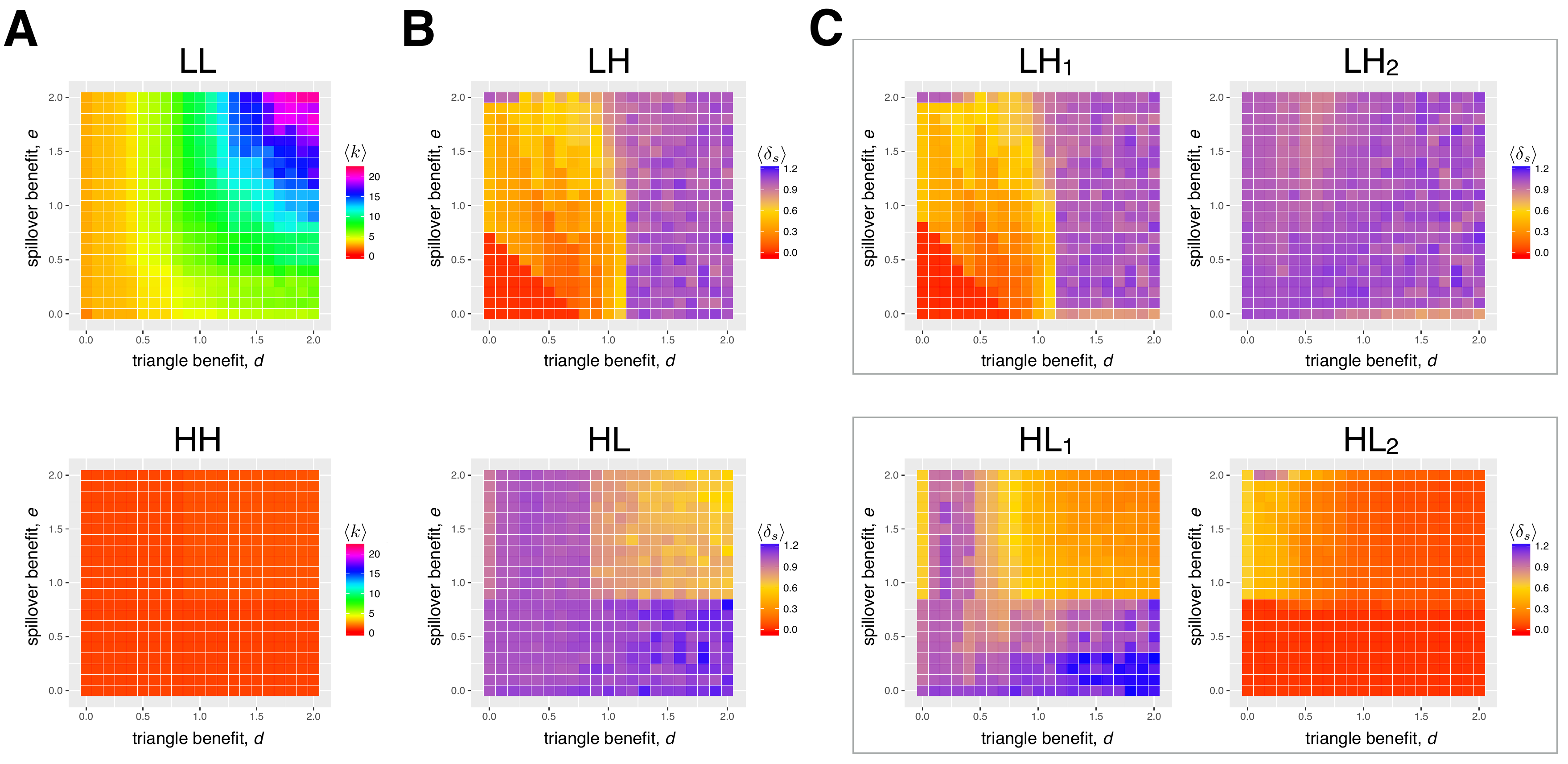}
\caption{\label{fig:fig7}  Mixed effects. (A) Average degree for the LL and HH (no-shock) conditions. (B) Resilience for HL and LH shock conditions. (C) Resilience for shock conditions in which only Layer 1 experienced a shock, so that tie costs in Layer 2 remained as they were before the shock. Comparison for resilience is to Layer 1 of the network shown in subfigure (A).}
\end{figure*}

\begin{figure*}
\includegraphics[width=0.7\textwidth]{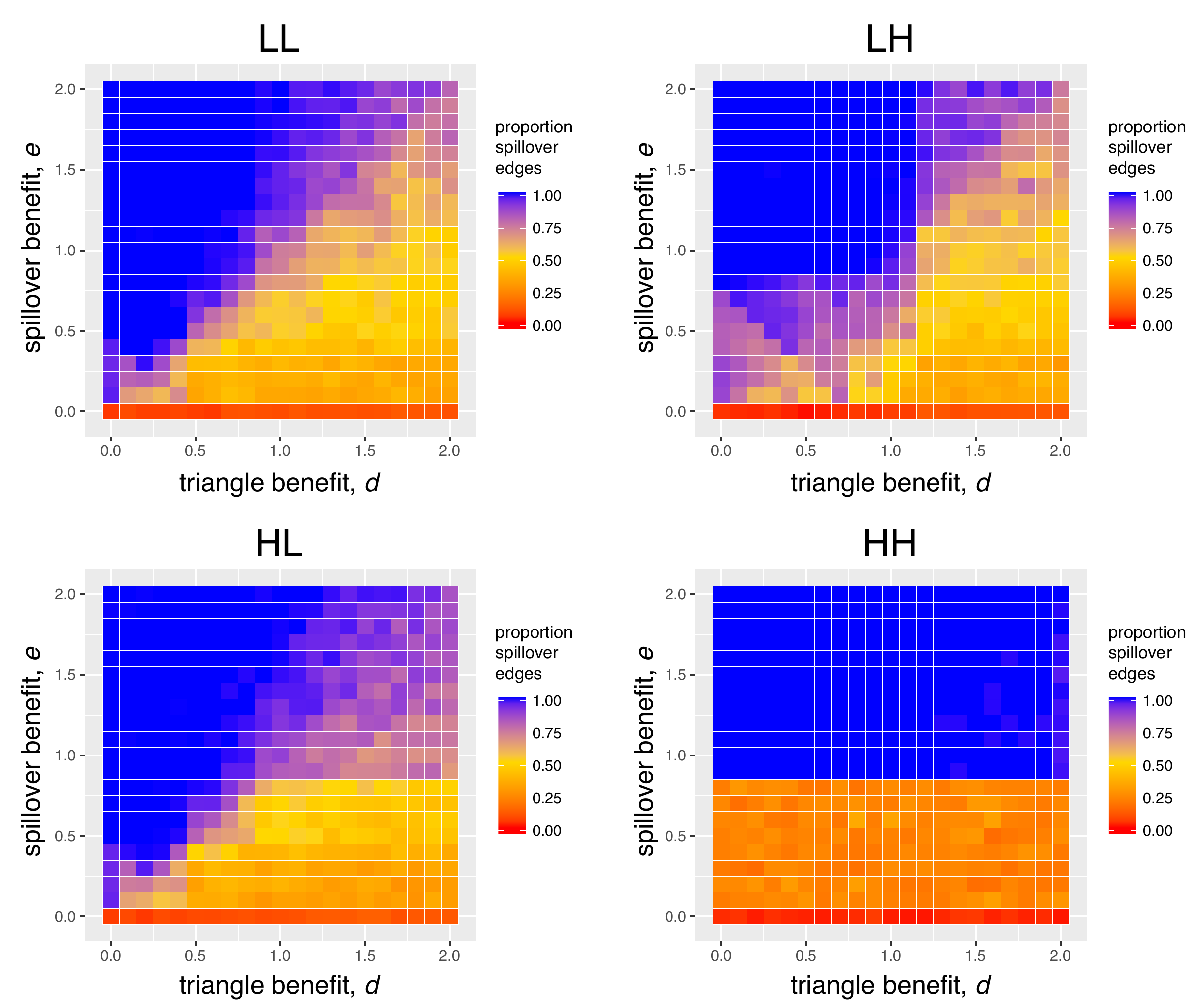}
\caption{\label{fig:fig8}  Heatmaps showing the proportion of all ties that are spillover under all four shock conditions. These data are averaged from 10 runs for each parameter condition.}
\end{figure*}

We can again quantify the emergent network structure by using average degree and resilience. Fig. \ref{fig:fig7}A shows the average degree at equilibrium for cases in which no shock occurred. Under high costs, average degree is largely unaffected by structural incentives. Supporting the visual inspection (see Fig. \ref{fig:fig6} ), we see that under low tie costs, the effects are largely additive, yielding quite dense networks when strong incentives for both closed triangles and spillover ties are present. 

Fig. \ref{fig:fig7}B shows the resilience for both shock conditions. 
Our main result is captured by the LH condition (and applies as well to the isolated effects cases):
structural incentives create resiliency in networks, and allow the retention of structural complexity after an increase in tie costs. This complexity could not possibly arise if costs were very high to begin with. 
As Fig. \ref{fig:fig2}A documents, with sufficiently high incentives for triangle closure, the average degree of ``shocked" (Low-High, and High-Low) network matches closely those of the unshocked (Low-Low) networks.
We also find that as long as the triangle benefit is high enough, additional spillover benefits do not influence average degree, though they do influence other aspects of network structure (Fig. \ref{fig:fig6}). For low values of $d$, stronger incentives for spillover ties can compensate to create resiliency, and similarly for low values of $e$ and incentives for closed triangles. Indeed, the effects of these incentives are additive in the model: resilience under mixed effects occurs whenever the sum of $d + e$ exceeds a critical threshold (see SI Appendix for derivation). 

For HL shocks, there is a regime under high structural benefits in which  resilience is actually lower than under smaller benefits (Fig. \ref{fig:fig7}B, bottom row). That is, when tie costs are initially high, the presence of strong structural incentives prevents the network from becoming as dense as it would have otherwise been once tie costs are lowered, compared with the density of comparable network in which initial tie costs are low. To explain this, observe that, in the regime of $e > 0.8$, all ties will be spillover ties (see Fig. \ref{fig:fig8}). In the regime of $d, e > 0.8$, the large structural incentives mean that any time an agent has the opportunity to 
add a new tie that either closes a triangle or completes a spillover at the expense of a current tie that does not do those things, it will do so. This also means that networks can become highly clustered across both layers, including the formation of two-layer triangles, even while tie costs are still high. An illustration of this is given in the SI Appendix (see Fig. S8). This in turn means that, for a given node, the opportunities for forming new ties when costs are lowered will be more constrained, and as such the emergent network structures may have fewer overall connections than if the network had been initialized with low tie costs. In contrast, when ties are consistently low, more two-stars will form by chance before their triangles are closed, leading to higher overall degree.

For most of the results presented thus far, data from only one layer of the two-layer multiplex was presented. This is justified because the initialization conditions and structural incentives are identical for each layer, so they will be statistically identical in structure. In terms of the overall network structure, the additional features involving spillover ties are not present. Fig. \ref{fig:fig8} shows the proportion of all ties that are spillover ties as a function of the structural incentives. This is calculated by doubling the number of spillover ties and then dividing by the sum of the total number of ties in each layer. We see that, for low costs, most if not all edges will be spillover ties whenever spillover benefit is strong relative to triangle benefit. Otherwise, about half of all ties are spillover edges. In the case of high tie costs, spillover ties were relatively rare until $e > 0.8$, which is when the benefit of a spillover tie became strictly larger than the utility a single, non-spillover tie. Therefore, a node that could drop its current tie in favor of a new one involving a spillover edge was suddenly incentivized. See SI Appendix for further discussion of spillover and network structure. 

\subsection{Mixed effects with shocks in only one layer}

In all cases presented so far, we assumed that shocks occurred 
in both layers of the multiplex. In some cases, however, changes to structural incentives might occur in only one layer. 
However, what happens in one social context can influence social behaviors in other contexts \cite{lubell2013ecology, brummitt2015coupled}.
To explore this idea with our model, we ran simulations in which the shock occurred only in one layer (always Layer 1). In other words, for LH (HL) shocks, both layers began with low (high) tie costs. After an initial equilibrium was reached, the cost of ties in Layer 1---but not Layer 2---increased (decreased). 

For LH shocks---that is, in cases where tie costs increased---results were largely unaffected by spillover (Fig.  \ref{fig:fig7}C, top row). The un-shocked layer was not different in its average degree than either layer of the baseline network that received no shock (Fig. \ref{fig:fig7}A, top row). Therefore the resilience of the un-shocked layer was uniformly high regardless of the type of incentives at work. The shocked layer was similar to the baseline layer that did receive a shock. (Fig. \ref{fig:fig7}B, top row). 

More interesting is the case of the HL shock, in which tie costs decreased from high to low (Fig. \ref{fig:fig7}C, bottom row). For low values of $e$, the benefit to spillover, each layer resembled the non-shocked network with the corresponding final cost. When $e > 0.8$, however,  we did observe a spillover effect in which the shocked layer grew more similar to the unshocked layer and vice versa, relative to baseline. This is because under such structural incentives and high tie costs, all ties are spillover ties (Fig. \ref{fig:fig8}). This means that all new edges formed initially after the shock will exist in one layer only, creating more opportunities to complete the spillover tie in the corresponding layer. The addition of new non-spillover links in the shocked layer after the shock creates incentives for some agents to then add a corresponding tie in the un-shocked layer, once the payoff for doing do outweighs the cost of adding a new edge. In contrast, there will be no spontaneous edge formation in the unshocked layer, and so the amount of new ties in the shocked layer will be diminished relative to the baseline case in which both layers are shocked.

\subsection{Lower costs}
To maximize clarity and illustrate stark differences between conditions, we purposefully chose a value for high tie costs ($c_{high}$) that would minimize emergent network structure. In this condition, triangles never form, because forming a triangle requires closing a two-star,  the formation of which is never incentivized with such high costs. Agents in our model are unable to sacrifice short-term costs for long-term gain. Very short chains of three nodes do sometimes form under large incentives for spillover tie, because the cost of   one additional tie can be overcome by the added benefit of the spillover tie it forms. This condition permits a clear contrast with our low-cost scenario, which does permit the formation of triangles and long chains. As such, we can show exactly the extent to which structural incentives provide resilience to the network once tie costs become high after a shock. 

That being said, it is important to illustrate that our main effect is robust to cases where {\em some} complex network structures emerge even under high costs. To test this, we repeated our mixed effects simulations (in which shocks affected both layers of the multiplex), but with $c_{high} = 0.3$. In this case, triangles and chains formed even under high costs, though the average degree of the network was still lower than under low costs (still set at $c_{low} = 0.2$). These results are described in the SI Appendix (see Fig. S11). We find that, although the zone of intermediate resiliency (in which $\delta_s$ is less than one but greater than zero) is much smaller than it is when $c_{high} = 0.6$, there are still clear zones of zero resilience when $d$ and $e$ are small and total resilience when they are large.

\subsection{Noise}
Our main results concerning resilience are entirely a consequence of the stable nature of structural features like triangles and spillover edges. Our model assumes that individuals will always act to maximize their utility and will never take an action that decreases their utility. 
The path-dependent nature of structural features induces (sometimes high) levels of resilience in networks undergoing shocks.  Under conditions of deterministic rationality and path-dependence, 
many social ties can remain incentivized even when tie costs increase. However, if one starts with an empty network, it is very much a case of ``you can't get there from here," to borrow a favorite New England colloquialism. 

This matters, because if noise or poor decision making were to destroy those incentivized structural features of the network, it is doubtful that they could be recovered. Under high tie costs, when noise leads to the suboptimal adding or dropping of ties, it is only the HH equilibria that are truly stable. The equilibria that result from LH shocks, which we have shown to demonstrate resilience, are only metastable.  An important question then becomes: how susceptible to noise are those LH equilibria (really pseudoequilibria under noise). This can be posed by asking how quickly following a shock, in which tie costs increase, does a network initialized with low costs revert to the state of network initialized with high costs. We find that, for a noise rate $\nu$, the number of rounds required for this to occur was approximately $1/\nu$ (see SI Appendix and Fig. S8). This is reassuring. If decisions are often made randomly or deviate significantly from optimality, path dependence will not matter a great deal to network formation. However, as long as levels of noise are reasonably low, path-dependency appears to have a substantial effect on network resilience. Specifically, with low levels of noise, it takes a network a long time to lose its pre-shock structural features.

\section{Discussion}

We see that quite interesting and varied networks can form under combined incentives for social ties, triangles, and spillover ties. More importantly, we have shown that 
when there exist sufficiently high incentives for closed triangles, spillover, or both, networks can be quite resilient to shocks due to structural entrenchment. 

Our results indicate how network structures can emerge and respond differently to local incentives for the formation and dissolution of social ties at different points in the network formation process. Moreover, the existence of spillover effects between layers of multiplex networks shows that historical events in one layer can change the structure of other layers. Our investigation therefore has broad implications for understanding the formation and evolution of complex social networks in many real world contexts. 
In particular, it highlights processes that induce dramatic changes in individual incentives, as well as processes that involve spillover between different social contexts.

Further extensions of our model may be useful for understanding a wider range of behaviors on multiplex networks in relation to shocks and resilience. For example, the addition of node-level heterogeneity would allow for the inclusion of homophilic or parochial behavior.  Heterogeneity could also be applied to  incentives, allowing different subsets of the population or different social context (corresponding to different layers of the multiplex) to vary in systematic and realistic ways. We assumed in the present analysis that all nodes in a layer experienced shocks at the same time. Exploring the implications of shocks affecting only a subset of nodes will be important for understanding how local events can propagate influence throughout a multiplex. In addition, we have only analyzed relatively small networks of 80 nodes or less. Although our analysis indicates that our results are likely to hold for networks within an order of  magnitude or more in size, exploration of the dynamics of substantially larger networks may be useful to test of the boundaries of our findings. 

Most real world systems are quite a bit more complex than the system expressed in our model. However, understanding the nature of resilience through structural entrenchment, as well as the influence of spillover in multiplex networks, may help guide both the analysis and collection of social network data in a wide variety of domains, from international relations to internet applications to the study of social ties across range of human cultures. Moreover, one can always add more complexity to a model in the name of increased realism. However, simple models such as ours, which are easier to understand and analyze, can nevertheless yield important insights of their own, and also provide a baseline from which to perform richer explorations \cite{Wimsatt:1987aa}. In this case, such parsimony allowed for the discovery and exploration of a novel network phenomenon.

\begin{acknowledgments}
[blinded]
We thank our colleagues on the SPINS project at UC Davis and the Naval Postgraduate School for comments, and we particularly thank Pierre-Andr\'{e} No\"{e}l, Camber Warren, and Haochen Wu for feedback. We are also grateful to an anonymous referee whose comments significantly strengthened the manuscript, in particular, providing a template for what became Fig S8 in the SI Appendix. 
This project was supported by a Minerva grant from the U. S. Department of Defense, \#W911NF-15-1-0502.
\end{acknowledgments}

\bibliography{multiplex}{}

\begin{thebibliography}{42}%
\makeatletter
\providecommand \@ifxundefined [1]{%
 \@ifx{#1\undefined}
}%
\providecommand \@ifnum [1]{%
 \ifnum #1\expandafter \@firstoftwo
 \else \expandafter \@secondoftwo
 \fi
}%
\providecommand \@ifx [1]{%
 \ifx #1\expandafter \@firstoftwo
 \else \expandafter \@secondoftwo
 \fi
}%
\providecommand \natexlab [1]{#1}%
\providecommand \enquote  [1]{``#1''}%
\providecommand \bibnamefont  [1]{#1}%
\providecommand \bibfnamefont [1]{#1}%
\providecommand \citenamefont [1]{#1}%
\providecommand \href@noop [0]{\@secondoftwo}%
\providecommand \href [0]{\begingroup \@sanitize@url \@href}%
\providecommand \@href[1]{\@@startlink{#1}\@@href}%
\providecommand \@@href[1]{\endgroup#1\@@endlink}%
\providecommand \@sanitize@url [0]{\catcode `\\12\catcode `\$12\catcode
  `\&12\catcode `\#12\catcode `\^12\catcode `\_12\catcode `\%12\relax}%
\providecommand \@@startlink[1]{}%
\providecommand \@@endlink[0]{}%
\providecommand \url  [0]{\begingroup\@sanitize@url \@url }%
\providecommand \@url [1]{\endgroup\@href {#1}{\urlprefix }}%
\providecommand \urlprefix  [0]{URL }%
\providecommand \Eprint [0]{\href }%
\providecommand \doibase [0]{http://dx.doi.org/}%
\providecommand \selectlanguage [0]{\@gobble}%
\providecommand \bibinfo  [0]{\@secondoftwo}%
\providecommand \bibfield  [0]{\@secondoftwo}%
\providecommand \translation [1]{[#1]}%
\providecommand \BibitemOpen [0]{}%
\providecommand \bibitemStop [0]{}%
\providecommand \bibitemNoStop [0]{.\EOS\space}%
\providecommand \EOS [0]{\spacefactor3000\relax}%
\providecommand \BibitemShut  [1]{\csname bibitem#1\endcsname}%
\let\auto@bib@innerbib\@empty
\bibitem [{\citenamefont {Barab\'{a}si}\ and\ \citenamefont
  {Albert}(1999)}]{barabasi1999scaling}%
  \BibitemOpen
  \bibfield  {author} {\bibinfo {author} {\bibfnamefont {A.-L.}\ \bibnamefont
  {Barab\'{a}si}}\ and\ \bibinfo {author} {\bibfnamefont {R.}~\bibnamefont
  {Albert}},\ }\href@noop {} {\bibfield  {journal} {\bibinfo  {journal}
  {Science}\ }\textbf {\bibinfo {volume} {286}},\ \bibinfo {pages} {509}
  (\bibinfo {year} {1999})}\BibitemShut {NoStop}%
\bibitem [{\citenamefont {Granovetter}(1973)}]{granovetter_weakties_1973}%
  \BibitemOpen
  \bibfield  {author} {\bibinfo {author} {\bibfnamefont {M.~S.}\ \bibnamefont
  {Granovetter}},\ }\href@noop {} {\bibfield  {journal} {\bibinfo  {journal}
  {American Journal of Sociology}\ }\textbf {\bibinfo {volume} {78}},\ \bibinfo
  {pages} {1360} (\bibinfo {year} {1973})}\BibitemShut {NoStop}%
\bibitem [{\citenamefont {Seeman}(1996)}]{seeman1996social}%
  \BibitemOpen
  \bibfield  {author} {\bibinfo {author} {\bibfnamefont {T.~E.}\ \bibnamefont
  {Seeman}},\ }\href@noop {} {\bibfield  {journal} {\bibinfo  {journal} {Annals
  of Epidemiology}\ }\textbf {\bibinfo {volume} {6}},\ \bibinfo {pages} {442}
  (\bibinfo {year} {1996})}\BibitemShut {NoStop}%
\bibitem [{\citenamefont {{Holt-Lunstad}}\ \emph {et~al.}(2010)\citenamefont
  {{Holt-Lunstad}}, \citenamefont {Smith},\ and\ \citenamefont
  {Layton}}]{holt2010social}%
  \BibitemOpen
  \bibfield  {author} {\bibinfo {author} {\bibfnamefont {J.}~\bibnamefont
  {{Holt-Lunstad}}}, \bibinfo {author} {\bibfnamefont {T.~B.}\ \bibnamefont
  {Smith}}, \ and\ \bibinfo {author} {\bibfnamefont {J.~B.}\ \bibnamefont
  {Layton}},\ }\href@noop {} {\bibfield  {journal} {\bibinfo  {journal} {PLOS
  Medicine}\ }\textbf {\bibinfo {volume} {7}},\ \bibinfo {pages} {e1000316}
  (\bibinfo {year} {2010})}\BibitemShut {NoStop}%
\bibitem [{\citenamefont {Cacioppo}\ and\ \citenamefont
  {Hawkley}(2009)}]{cacioppo2009perceived}%
  \BibitemOpen
  \bibfield  {author} {\bibinfo {author} {\bibfnamefont {J.~T.}\ \bibnamefont
  {Cacioppo}}\ and\ \bibinfo {author} {\bibfnamefont {L.~C.}\ \bibnamefont
  {Hawkley}},\ }\href@noop {} {\bibfield  {journal} {\bibinfo  {journal}
  {Trends in Cognitive Science}\ }\textbf {\bibinfo {volume} {13}},\ \bibinfo
  {pages} {447} (\bibinfo {year} {2009})}\BibitemShut {NoStop}%
\bibitem [{\citenamefont {Coleman}(1988)}]{coleman1988capital}%
  \BibitemOpen
  \bibfield  {author} {\bibinfo {author} {\bibfnamefont {J.~S.}\ \bibnamefont
  {Coleman}},\ }\href@noop {} {\bibfield  {journal} {\bibinfo  {journal}
  {American Journal of Sociology}\ }\textbf {\bibinfo {volume} {94}},\ \bibinfo
  {pages} {S95} (\bibinfo {year} {1988})}\BibitemShut {NoStop}%
\bibitem [{\citenamefont {Coleman}(1990)}]{coleman1990social}%
  \BibitemOpen
  \bibfield  {author} {\bibinfo {author} {\bibfnamefont {J.~S.}\ \bibnamefont
  {Coleman}},\ }\href@noop {} {\emph {\bibinfo {title} {Foundations of a social
  theory}}}\ (\bibinfo  {publisher} {Harvard University Press},\ \bibinfo
  {address} {Cambridge, MA},\ \bibinfo {year} {1990})\BibitemShut {NoStop}%
\bibitem [{\citenamefont {Long}(1958)}]{long_local_1958}%
  \BibitemOpen
  \bibfield  {author} {\bibinfo {author} {\bibfnamefont {N.~E.}\ \bibnamefont
  {Long}},\ }\href@noop {} {\bibfield  {journal} {\bibinfo  {journal} {American
  Journal of Sociology}\ }\textbf {\bibinfo {volume} {64}},\ \bibinfo {pages}
  {251} (\bibinfo {year} {1958})}\BibitemShut {NoStop}%
\bibitem [{\citenamefont {Hinde}(1976)}]{hinde1976inter}%
  \BibitemOpen
  \bibfield  {author} {\bibinfo {author} {\bibfnamefont {R.~A.}\ \bibnamefont
  {Hinde}},\ }\href@noop {} {\bibfield  {journal} {\bibinfo  {journal} {Man}\
  }\textbf {\bibinfo {volume} {11}},\ \bibinfo {pages} {1} (\bibinfo {year}
  {1976})}\BibitemShut {NoStop}%
\bibitem [{\citenamefont {Putnam}(2000)}]{putnam2000bowling}%
  \BibitemOpen
  \bibfield  {author} {\bibinfo {author} {\bibfnamefont {R.~D.}\ \bibnamefont
  {Putnam}},\ }\href@noop {} {\emph {\bibinfo {title} {Bowling alone: The
  collapse and revival of {A}merican community}}}\ (\bibinfo  {publisher}
  {Simon \& Schuster},\ \bibinfo {address} {New York},\ \bibinfo {year}
  {2000})\BibitemShut {NoStop}%
\bibitem [{\citenamefont {Ashmore}\ \emph {et~al.}(2004)\citenamefont
  {Ashmore}, \citenamefont {Deaux},\ and\ \citenamefont
  {{McLaughlin-Volpe}}}]{ashmore2004organizing}%
  \BibitemOpen
  \bibfield  {author} {\bibinfo {author} {\bibfnamefont {R.~D.}\ \bibnamefont
  {Ashmore}}, \bibinfo {author} {\bibfnamefont {K.}~\bibnamefont {Deaux}}, \
  and\ \bibinfo {author} {\bibfnamefont {T.}~\bibnamefont
  {{McLaughlin-Volpe}}},\ }\href@noop {} {\bibfield  {journal} {\bibinfo
  {journal} {Psychological Bulletin}\ }\textbf {\bibinfo {volume} {130}},\
  \bibinfo {pages} {80} (\bibinfo {year} {2004})}\BibitemShut {NoStop}%
\bibitem [{\citenamefont {Smaldino}(2017)}]{smaldinoInPressID}%
  \BibitemOpen
  \bibfield  {author} {\bibinfo {author} {\bibfnamefont {P.~E.}\ \bibnamefont
  {Smaldino}},\ }in\ \href@noop {} {\emph {\bibinfo {booktitle} {Beyond the
  Meme: Dynamical Structures in Cultural Evolution}}},\ \bibinfo {editor}
  {edited by\ \bibinfo {editor} {\bibfnamefont {A.~C.}\ \bibnamefont {Love}}\
  and\ \bibinfo {editor} {\bibfnamefont {W.~C.}\ \bibnamefont {Wimsatt}}}\
  (\bibinfo  {publisher} {Univ Minnesota Press},\ \bibinfo {year}
  {{2017}})\BibitemShut {NoStop}%
\bibitem [{\citenamefont {Wimsatt}(1987)}]{Wimsatt:1987aa}%
  \BibitemOpen
  \bibfield  {author} {\bibinfo {author} {\bibfnamefont {W.~C.}\ \bibnamefont
  {Wimsatt}},\ }in\ \href@noop {} {\emph {\bibinfo {booktitle} {Neutral Models
  in Biology}}},\ \bibinfo {editor} {edited by\ \bibinfo {editor}
  {\bibfnamefont {M.}~\bibnamefont {Nitecki}}\ and\ \bibinfo {editor}
  {\bibfnamefont {A.}~\bibnamefont {Hoffman}}}\ (\bibinfo  {publisher} {Oxford
  University Press},\ \bibinfo {address} {London},\ \bibinfo {year} {1987})\
  pp.\ \bibinfo {pages} {23--55}\BibitemShut {NoStop}%
\bibitem [{\citenamefont {Epstein}(2008)}]{epstein08}%
  \BibitemOpen
  \bibfield  {author} {\bibinfo {author} {\bibfnamefont {J.~M.}\ \bibnamefont
  {Epstein}},\ }\href@noop {} {\bibfield  {journal} {\bibinfo  {journal} {J.
  Artif. Soc. Soc. Simul.}\ }\textbf {\bibinfo {volume} {11}},\ \bibinfo
  {pages} {12} (\bibinfo {year} {2008})}\BibitemShut {NoStop}%
\bibitem [{\citenamefont {Smaldino}(2016)}]{smaldinoModels}%
  \BibitemOpen
  \bibfield  {author} {\bibinfo {author} {\bibfnamefont {P.~E.}\ \bibnamefont
  {Smaldino}},\ }in\ \href@noop {} {\emph {\bibinfo {booktitle} {Computational
  models in social psychology}}},\ \bibinfo {editor} {edited by\ \bibinfo
  {editor} {\bibfnamefont {R.~R.}\ \bibnamefont {Vallacher}}, \bibinfo {editor}
  {\bibfnamefont {A.}~\bibnamefont {Nowak}}, \ and\ \bibinfo {editor}
  {\bibfnamefont {S.~J.}\ \bibnamefont {Read}}}\ (\bibinfo  {publisher}
  {Psychology Press},\ \bibinfo {year} {{2016}})\BibitemShut {NoStop}%
\bibitem [{\citenamefont {Apicella}\ \emph {et~al.}(2012)\citenamefont
  {Apicella}, \citenamefont {Marlowe}, \citenamefont {Fowler},\ and\
  \citenamefont {Christakis}}]{apicella2012social}%
  \BibitemOpen
  \bibfield  {author} {\bibinfo {author} {\bibfnamefont {C.~L.}\ \bibnamefont
  {Apicella}}, \bibinfo {author} {\bibfnamefont {F.~W.}\ \bibnamefont
  {Marlowe}}, \bibinfo {author} {\bibfnamefont {J.~H.}\ \bibnamefont {Fowler}},
  \ and\ \bibinfo {author} {\bibfnamefont {N.~A.}\ \bibnamefont {Christakis}},\
  }\href@noop {} {\bibfield  {journal} {\bibinfo  {journal} {Nature}\ }\textbf
  {\bibinfo {volume} {481}},\ \bibinfo {pages} {497} (\bibinfo {year}
  {2012})}\BibitemShut {NoStop}%
\bibitem [{\citenamefont {Cohen}\ and\ \citenamefont
  {Haun}(2013)}]{cohen2013tags}%
  \BibitemOpen
  \bibfield  {author} {\bibinfo {author} {\bibfnamefont {E.}~\bibnamefont
  {Cohen}}\ and\ \bibinfo {author} {\bibfnamefont {D.}~\bibnamefont {Haun}},\
  }\href@noop {} {\bibfield  {journal} {\bibinfo  {journal} {Evolution and
  Human Behavior}\ }\textbf {\bibinfo {volume} {34}},\ \bibinfo {pages} {230}
  (\bibinfo {year} {2013})}\BibitemShut {NoStop}%
\bibitem [{\citenamefont {Lazer}\ and\ \citenamefont
  {Friedman}(2007)}]{lazer2007networks}%
  \BibitemOpen
  \bibfield  {author} {\bibinfo {author} {\bibfnamefont {D.}~\bibnamefont
  {Lazer}}\ and\ \bibinfo {author} {\bibfnamefont {A.}~\bibnamefont
  {Friedman}},\ }\href@noop {} {\bibfield  {journal} {\bibinfo  {journal}
  {Administrative Science Quarterly}\ }\textbf {\bibinfo {volume} {52}},\
  \bibinfo {pages} {667} (\bibinfo {year} {2007})}\BibitemShut {NoStop}%
\bibitem [{\citenamefont {Derex}\ and\ \citenamefont
  {Boyd}(2015)}]{derex2015foundations}%
  \BibitemOpen
  \bibfield  {author} {\bibinfo {author} {\bibfnamefont {M.}~\bibnamefont
  {Derex}}\ and\ \bibinfo {author} {\bibfnamefont {R.}~\bibnamefont {Boyd}},\
  }\href@noop {} {\bibfield  {journal} {\bibinfo  {journal} {Nature
  Communications}\ }\textbf {\bibinfo {volume} {6}},\ \bibinfo {pages} {8398}
  (\bibinfo {year} {2015})}\BibitemShut {NoStop}%
\bibitem [{\citenamefont {Centola}(2015)}]{centola2015social}%
  \BibitemOpen
  \bibfield  {author} {\bibinfo {author} {\bibfnamefont {D.}~\bibnamefont
  {Centola}},\ }\href@noop {} {\bibfield  {journal} {\bibinfo  {journal}
  {American Journal of Sociology}\ }\textbf {\bibinfo {volume} {120}},\
  \bibinfo {pages} {1295} (\bibinfo {year} {2015})}\BibitemShut {NoStop}%
\bibitem [{\citenamefont {Jackson}\ and\ \citenamefont
  {Watts}(2002)}]{jackson2002evolution}%
  \BibitemOpen
  \bibfield  {author} {\bibinfo {author} {\bibfnamefont {M.~O.}\ \bibnamefont
  {Jackson}}\ and\ \bibinfo {author} {\bibfnamefont {A.}~\bibnamefont
  {Watts}},\ }\href@noop {} {\bibfield  {journal} {\bibinfo  {journal} {Journal
  of Economic Theory}\ }\textbf {\bibinfo {volume} {106}},\ \bibinfo {pages}
  {265} (\bibinfo {year} {2002})}\BibitemShut {NoStop}%
\bibitem [{\citenamefont {Schweitzer}\ \emph {et~al.}(2009)\citenamefont
  {Schweitzer}, \citenamefont {Fagiolo}, \citenamefont {Sornette},
  \citenamefont {{Vega-Redondo}}, \citenamefont {Vespignano},\ and\
  \citenamefont {White}}]{schweitzer2009economic}%
  \BibitemOpen
  \bibfield  {author} {\bibinfo {author} {\bibfnamefont {F.}~\bibnamefont
  {Schweitzer}}, \bibinfo {author} {\bibfnamefont {G.}~\bibnamefont {Fagiolo}},
  \bibinfo {author} {\bibfnamefont {D.}~\bibnamefont {Sornette}}, \bibinfo
  {author} {\bibfnamefont {F.}~\bibnamefont {{Vega-Redondo}}}, \bibinfo
  {author} {\bibfnamefont {A.}~\bibnamefont {Vespignano}}, \ and\ \bibinfo
  {author} {\bibfnamefont {D.~R.}\ \bibnamefont {White}},\ }\href@noop {}
  {\bibfield  {journal} {\bibinfo  {journal} {Science}\ }\textbf {\bibinfo
  {volume} {325}},\ \bibinfo {pages} {422} (\bibinfo {year}
  {2009})}\BibitemShut {NoStop}%
\bibitem [{\citenamefont {Saram\"{a}ki}\ \emph {et~al.}(2014)\citenamefont
  {Saram\"{a}ki}, \citenamefont {Leicht}, \citenamefont {L\'{o}pez},
  \citenamefont {Roberts}, \citenamefont {{Reed-Tsochas}},\ and\ \citenamefont
  {Dunbar}}]{saramaki2014}%
  \BibitemOpen
  \bibfield  {author} {\bibinfo {author} {\bibfnamefont {J.}~\bibnamefont
  {Saram\"{a}ki}}, \bibinfo {author} {\bibfnamefont {E.~A.}\ \bibnamefont
  {Leicht}}, \bibinfo {author} {\bibfnamefont {E.}~\bibnamefont {L\'{o}pez}},
  \bibinfo {author} {\bibfnamefont {S.~G.~B.}\ \bibnamefont {Roberts}},
  \bibinfo {author} {\bibfnamefont {F.}~\bibnamefont {{Reed-Tsochas}}}, \ and\
  \bibinfo {author} {\bibfnamefont {R.~I.~M.}\ \bibnamefont {Dunbar}},\
  }\href@noop {} {\bibfield  {journal} {\bibinfo  {journal} {Proc. Natl. Acad.
  Sci. USA}\ }\textbf {\bibinfo {volume} {111}},\ \bibinfo {pages} {942}
  (\bibinfo {year} {2014})}\BibitemShut {NoStop}%
\bibitem [{\citenamefont {Palla}\ \emph {et~al.}(2005)\citenamefont {Palla},
  \citenamefont {Der{\'e}nyi}, \citenamefont {Farkas},\ and\ \citenamefont
  {Vicsek}}]{palla2005}%
  \BibitemOpen
  \bibfield  {author} {\bibinfo {author} {\bibfnamefont {G.}~\bibnamefont
  {Palla}}, \bibinfo {author} {\bibfnamefont {I.}~\bibnamefont {Der{\'e}nyi}},
  \bibinfo {author} {\bibfnamefont {I.}~\bibnamefont {Farkas}}, \ and\ \bibinfo
  {author} {\bibfnamefont {T.}~\bibnamefont {Vicsek}},\ }\href@noop {}
  {\bibfield  {journal} {\bibinfo  {journal} {Nature}\ }\textbf {\bibinfo
  {volume} {435}},\ \bibinfo {pages} {814} (\bibinfo {year}
  {2005})}\BibitemShut {NoStop}%
\bibitem [{\citenamefont {Cai}\ \emph {et~al.}(2005)\citenamefont {Cai},
  \citenamefont {Shao}, \citenamefont {He}, \citenamefont {Yan},\ and\
  \citenamefont {Han}}]{cai2005}%
  \BibitemOpen
  \bibfield  {author} {\bibinfo {author} {\bibfnamefont {D.}~\bibnamefont
  {Cai}}, \bibinfo {author} {\bibfnamefont {Z.}~\bibnamefont {Shao}}, \bibinfo
  {author} {\bibfnamefont {X.}~\bibnamefont {He}}, \bibinfo {author}
  {\bibfnamefont {X.}~\bibnamefont {Yan}}, \ and\ \bibinfo {author}
  {\bibfnamefont {J.}~\bibnamefont {Han}},\ }in\ \href@noop {} {\emph {\bibinfo
  {booktitle} {European Conference on Principles of Data Mining and Knowledge
  Discovery}}}\ (\bibinfo {organization} {Springer},\ \bibinfo {year} {2005})\
  pp.\ \bibinfo {pages} {445--452}\BibitemShut {NoStop}%
\bibitem [{\citenamefont {Lubell}(2013)}]{lubell2013ecology}%
  \BibitemOpen
  \bibfield  {author} {\bibinfo {author} {\bibfnamefont {M.}~\bibnamefont
  {Lubell}},\ }\href@noop {} {\bibfield  {journal} {\bibinfo  {journal} {Policy
  Studies Journal}\ }\textbf {\bibinfo {volume} {41}},\ \bibinfo {pages} {537}
  (\bibinfo {year} {2013})}\BibitemShut {NoStop}%
\bibitem [{\citenamefont {Vijayaraghavan}\ \emph {et~al.}(2015)\citenamefont
  {Vijayaraghavan}, \citenamefont {No{\"{e}l}}, \citenamefont {Maoz},\ and\
  \citenamefont {{D'Souza}}}]{vijayaraghavan2015spillover}%
  \BibitemOpen
  \bibfield  {author} {\bibinfo {author} {\bibfnamefont {V.~S.}\ \bibnamefont
  {Vijayaraghavan}}, \bibinfo {author} {\bibfnamefont {P.-A.}\ \bibnamefont
  {No{\"{e}l}}}, \bibinfo {author} {\bibfnamefont {Z.}~\bibnamefont {Maoz}}, \
  and\ \bibinfo {author} {\bibfnamefont {R.~M.}\ \bibnamefont {{D'Souza}}},\
  }\href@noop {} {\bibfield  {journal} {\bibinfo  {journal} {Scientific
  Reports}\ }\textbf {\bibinfo {volume} {5}},\ \bibinfo {pages} {15142}
  (\bibinfo {year} {2015})}\BibitemShut {NoStop}%
\bibitem [{\citenamefont {Brummitt}\ \emph {et~al.}(2015)\citenamefont
  {Brummitt}, \citenamefont {Barnett},\ and\ \citenamefont
  {{D'Souza}}}]{brummitt2015coupled}%
  \BibitemOpen
  \bibfield  {author} {\bibinfo {author} {\bibfnamefont {C.~D.}\ \bibnamefont
  {Brummitt}}, \bibinfo {author} {\bibfnamefont {G.}~\bibnamefont {Barnett}}, \
  and\ \bibinfo {author} {\bibfnamefont {R.~M.}\ \bibnamefont {{D'Souza}}},\
  }\href@noop {} {\bibfield  {journal} {\bibinfo  {journal} {J. R. Soc.
  Interface}\ }\textbf {\bibinfo {volume} {12}},\ \bibinfo {pages} {20150712}
  (\bibinfo {year} {2015})}\BibitemShut {NoStop}%
\bibitem [{\citenamefont {Kivel\"{a}}\ \emph {et~al.}(2014)\citenamefont
  {Kivel\"{a}}, \citenamefont {Arenas}, \citenamefont {Barthelemy},
  \citenamefont {Gleeson}, \citenamefont {Moreno},\ and\ \citenamefont
  {Porter}}]{kivela2014multilayer}%
  \BibitemOpen
  \bibfield  {author} {\bibinfo {author} {\bibfnamefont {M.}~\bibnamefont
  {Kivel\"{a}}}, \bibinfo {author} {\bibfnamefont {A.}~\bibnamefont {Arenas}},
  \bibinfo {author} {\bibfnamefont {M.}~\bibnamefont {Barthelemy}}, \bibinfo
  {author} {\bibfnamefont {J.~P.}\ \bibnamefont {Gleeson}}, \bibinfo {author}
  {\bibfnamefont {Y.}~\bibnamefont {Moreno}}, \ and\ \bibinfo {author}
  {\bibfnamefont {M.~A.}\ \bibnamefont {Porter}},\ }\href@noop {} {\bibfield
  {journal} {\bibinfo  {journal} {Journal of Complex Networks}\ }\textbf
  {\bibinfo {volume} {2}},\ \bibinfo {pages} {203} (\bibinfo {year}
  {2014})}\BibitemShut {NoStop}%
\bibitem [{\citenamefont {Boccaletti}\ \emph {et~al.}(2014)\citenamefont
  {Boccaletti}, \citenamefont {Bianconi}, \citenamefont {Criado}, \citenamefont
  {{del Genio}}, \citenamefont {{G\'{o}mez-Garde\~{n}es}}, \citenamefont
  {Romance}, \citenamefont {{Sendi\~{n}a-Nadal}}, \citenamefont {Wang},\ and\
  \citenamefont {Zanin}}]{boccaletti2014structure}%
  \BibitemOpen
  \bibfield  {author} {\bibinfo {author} {\bibfnamefont {S.}~\bibnamefont
  {Boccaletti}}, \bibinfo {author} {\bibfnamefont {G.}~\bibnamefont
  {Bianconi}}, \bibinfo {author} {\bibfnamefont {R.}~\bibnamefont {Criado}},
  \bibinfo {author} {\bibfnamefont {C.~I.}\ \bibnamefont {{del Genio}}},
  \bibinfo {author} {\bibfnamefont {J.}~\bibnamefont
  {{G\'{o}mez-Garde\~{n}es}}}, \bibinfo {author} {\bibfnamefont
  {M.}~\bibnamefont {Romance}}, \bibinfo {author} {\bibfnamefont
  {I.}~\bibnamefont {{Sendi\~{n}a-Nadal}}}, \bibinfo {author} {\bibfnamefont
  {Z.}~\bibnamefont {Wang}}, \ and\ \bibinfo {author} {\bibfnamefont
  {M.}~\bibnamefont {Zanin}},\ }\href@noop {} {\bibfield  {journal} {\bibinfo
  {journal} {Physics Reports}\ }\textbf {\bibinfo {volume} {544}},\ \bibinfo
  {pages} {1} (\bibinfo {year} {2014})}\BibitemShut {NoStop}%
\bibitem [{\citenamefont {Nicosia}\ \emph {et~al.}(2013)\citenamefont
  {Nicosia}, \citenamefont {Bianconi}, \citenamefont {Latora},\ and\
  \citenamefont {Barthelemy}}]{nicosia2013growing}%
  \BibitemOpen
  \bibfield  {author} {\bibinfo {author} {\bibfnamefont {V.}~\bibnamefont
  {Nicosia}}, \bibinfo {author} {\bibfnamefont {G.}~\bibnamefont {Bianconi}},
  \bibinfo {author} {\bibfnamefont {V.}~\bibnamefont {Latora}}, \ and\ \bibinfo
  {author} {\bibfnamefont {M.}~\bibnamefont {Barthelemy}},\ }\href@noop {}
  {\bibfield  {journal} {\bibinfo  {journal} {Physical Review Letters}\
  }\textbf {\bibinfo {volume} {111}},\ \bibinfo {pages} {058701} (\bibinfo
  {year} {2013})}\BibitemShut {NoStop}%
\bibitem [{\citenamefont {Kim}\ and\ \citenamefont
  {Goh}(2013)}]{kim2013coevolution}%
  \BibitemOpen
  \bibfield  {author} {\bibinfo {author} {\bibfnamefont {J.~Y.}\ \bibnamefont
  {Kim}}\ and\ \bibinfo {author} {\bibfnamefont {K.-I.}\ \bibnamefont {Goh}},\
  }\href@noop {} {\bibfield  {journal} {\bibinfo  {journal} {Physical Review
  Letters}\ }\textbf {\bibinfo {volume} {111}},\ \bibinfo {pages} {058702}
  (\bibinfo {year} {2013})}\BibitemShut {NoStop}%
\bibitem [{\citenamefont {Cardillo}\ \emph {et~al.}(2013)\citenamefont
  {Cardillo}, \citenamefont {{G\'{o}mez-Garde\~{n}es}}, \citenamefont {Zanin},
  \citenamefont {Romance}, \citenamefont {Papo}, \citenamefont {{del Pozo}},\
  and\ \citenamefont {Boccaletti}}]{cardillo2013emergence}%
  \BibitemOpen
  \bibfield  {author} {\bibinfo {author} {\bibfnamefont {A.}~\bibnamefont
  {Cardillo}}, \bibinfo {author} {\bibfnamefont {J.}~\bibnamefont
  {{G\'{o}mez-Garde\~{n}es}}}, \bibinfo {author} {\bibfnamefont
  {M.}~\bibnamefont {Zanin}}, \bibinfo {author} {\bibfnamefont
  {M.}~\bibnamefont {Romance}}, \bibinfo {author} {\bibfnamefont
  {D.}~\bibnamefont {Papo}}, \bibinfo {author} {\bibfnamefont {F.}~\bibnamefont
  {{del Pozo}}}, \ and\ \bibinfo {author} {\bibfnamefont {S.}~\bibnamefont
  {Boccaletti}},\ }\href@noop {} {\bibfield  {journal} {\bibinfo  {journal}
  {Scientific Reports}\ }\textbf {\bibinfo {volume} {3}},\ \bibinfo {pages}
  {1344} (\bibinfo {year} {2013})}\BibitemShut {NoStop}%
\bibitem [{\citenamefont {{G\'{o}mez-Garde\~{n}es}}\ \emph
  {et~al.}(2015)\citenamefont {{G\'{o}mez-Garde\~{n}es}}, \citenamefont {{de
  Domenico}}, \citenamefont {Guti\'{e}rrez}, \citenamefont {Arenas},\ and\
  \citenamefont {G\'{o}mez}}]{gomez2015layer}%
  \BibitemOpen
  \bibfield  {author} {\bibinfo {author} {\bibfnamefont {J.}~\bibnamefont
  {{G\'{o}mez-Garde\~{n}es}}}, \bibinfo {author} {\bibfnamefont
  {M.}~\bibnamefont {{de Domenico}}}, \bibinfo {author} {\bibfnamefont
  {G.}~\bibnamefont {Guti\'{e}rrez}}, \bibinfo {author} {\bibfnamefont
  {A.}~\bibnamefont {Arenas}}, \ and\ \bibinfo {author} {\bibfnamefont
  {S.}~\bibnamefont {G\'{o}mez}},\ }\href@noop {} {\bibfield  {journal}
  {\bibinfo  {journal} {Phil. Trans. R. Soc. A}\ }\textbf {\bibinfo {volume}
  {373}},\ \bibinfo {pages} {20150117.} (\bibinfo {year} {2015})}\BibitemShut
  {NoStop}%
\bibitem [{\citenamefont {Bianconi}(2013)}]{bianconi2013}%
  \BibitemOpen
  \bibfield  {author} {\bibinfo {author} {\bibfnamefont {G.}~\bibnamefont
  {Bianconi}},\ }\href@noop {} {\bibfield  {journal} {\bibinfo  {journal}
  {Phys. Rev. E}\ }\textbf {\bibinfo {volume} {87}},\ \bibinfo {pages} {062806}
  (\bibinfo {year} {2013})}\BibitemShut {NoStop}%
\bibitem [{\citenamefont {Batiston}\ \emph {et~al.}(2016)\citenamefont
  {Batiston}, \citenamefont {Iacovacci}, \citenamefont {Nicosia}, \citenamefont
  {Bianconi},\ and\ \citenamefont {Latora}}]{battiston2016}%
  \BibitemOpen
  \bibfield  {author} {\bibinfo {author} {\bibfnamefont {F.}~\bibnamefont
  {Batiston}}, \bibinfo {author} {\bibfnamefont {J.}~\bibnamefont {Iacovacci}},
  \bibinfo {author} {\bibfnamefont {V.}~\bibnamefont {Nicosia}}, \bibinfo
  {author} {\bibfnamefont {G.}~\bibnamefont {Bianconi}}, \ and\ \bibinfo
  {author} {\bibfnamefont {V.}~\bibnamefont {Latora}},\ }\href@noop {}
  {\bibfield  {journal} {\bibinfo  {journal} {PLOS ONE}\ }\textbf {\bibinfo
  {volume} {11}},\ \bibinfo {pages} {e0147451} (\bibinfo {year}
  {2016})}\BibitemShut {NoStop}%
\bibitem [{\citenamefont {Palchykov}\ \emph {et~al.}(2012)\citenamefont
  {Palchykov}, \citenamefont {Kaski}, \citenamefont {Kert\'{e}z}, \citenamefont
  {Barab\'{a}si},\ and\ \citenamefont {Dunbar}}]{palchykov2012}%
  \BibitemOpen
  \bibfield  {author} {\bibinfo {author} {\bibfnamefont {V.}~\bibnamefont
  {Palchykov}}, \bibinfo {author} {\bibfnamefont {K.}~\bibnamefont {Kaski}},
  \bibinfo {author} {\bibfnamefont {J.}~\bibnamefont {Kert\'{e}z}}, \bibinfo
  {author} {\bibfnamefont {A.-L.}\ \bibnamefont {Barab\'{a}si}}, \ and\
  \bibinfo {author} {\bibfnamefont {R.~I.~M.}\ \bibnamefont {Dunbar}},\
  }\href@noop {} {\bibfield  {journal} {\bibinfo  {journal} {Sci. Rep.}\
  }\textbf {\bibinfo {volume} {2}},\ \bibinfo {pages} {370} (\bibinfo {year}
  {2012})}\BibitemShut {NoStop}%
\bibitem [{\citenamefont {Hruschka}(2010)}]{hruschka2010}%
  \BibitemOpen
  \bibfield  {author} {\bibinfo {author} {\bibfnamefont {D.~J.}\ \bibnamefont
  {Hruschka}},\ }\href@noop {} {\emph {\bibinfo {title} {Friendship:
  {D}evelopment, ecology, and evolution of a relationship}}}\ (\bibinfo
  {publisher} {University of California Press},\ \bibinfo {address}
  {Berkeley},\ \bibinfo {year} {2010})\BibitemShut {NoStop}%
\bibitem [{\citenamefont {Burger}\ and\ \citenamefont
  {Buskens}(2009)}]{burger2009network}%
  \BibitemOpen
  \bibfield  {author} {\bibinfo {author} {\bibfnamefont {M.~J.}\ \bibnamefont
  {Burger}}\ and\ \bibinfo {author} {\bibfnamefont {V.}~\bibnamefont
  {Buskens}},\ }\href@noop {} {\bibfield  {journal} {\bibinfo  {journal}
  {Social Networks}\ }\textbf {\bibinfo {volume} {31}},\ \bibinfo {pages} {63}
  (\bibinfo {year} {2009})}\BibitemShut {NoStop}%
\bibitem [{\citenamefont {Simon}(1990)}]{simon1990invariants}%
  \BibitemOpen
  \bibfield  {author} {\bibinfo {author} {\bibfnamefont {H.~A.}\ \bibnamefont
  {Simon}},\ }\href@noop {} {\bibfield  {journal} {\bibinfo  {journal} {Annual
  Review of Psychology}\ }\textbf {\bibinfo {volume} {41}},\ \bibinfo {pages}
  {1} (\bibinfo {year} {1990})}\BibitemShut {NoStop}%
\bibitem [{\citenamefont {Smaldino}\ and\ \citenamefont
  {Lubell}(2011)}]{smaldino2011institutional}%
  \BibitemOpen
  \bibfield  {author} {\bibinfo {author} {\bibfnamefont {P.~E.}\ \bibnamefont
  {Smaldino}}\ and\ \bibinfo {author} {\bibfnamefont {M.}~\bibnamefont
  {Lubell}},\ }\href@noop {} {\bibfield  {journal} {\bibinfo  {journal} {PLOS
  ONE}\ }\textbf {\bibinfo {volume} {6}},\ \bibinfo {pages} {e23019} (\bibinfo
  {year} {2011})}\BibitemShut {NoStop}%
\bibitem [{\citenamefont {Burt}(1992)}]{burt1992structural}%
  \BibitemOpen
  \bibfield  {author} {\bibinfo {author} {\bibfnamefont {R.~S.}\ \bibnamefont
  {Burt}},\ }\href@noop {} {\emph {\bibinfo {title} {Structural holes: The
  social structure of competition}}}\ (\bibinfo  {publisher} {Harvard
  University Press},\ \bibinfo {address} {Cambridge, MA},\ \bibinfo {year}
  {1992})\BibitemShut {NoStop}%
\end{thebibliography}%

\newpage

\begin{center}
 {\bf
{\large SI APPENDIX\\~\large
(Online Supplement)
}
}
\end{center}
\appendix
\renewcommand\thefigure{\thesection.\arabic{figure}}    
\setcounter{figure}{0} 

\section{Exploration of Isolated Effects}
\subsection{Isolated effects: Triangle benefits only}

We first examine single-layer networks (for which no spillover is possible) in which there are additional benefits to closed triangles.  Network statistics are presented in Fig. \ref{fig:fig2}. 
Under low costs, many triangles form, and the average degree of the network increases as triangles are incentivized more (Fig. \ref{fig:fig2}A). This is because closed triangles scaffold the creation of addition triangles by providing affordances (e.g., new two-stars), forming a cascade. Such a cascade does not go on indefinitely, however. The costs of ties can set a practical limit, especially when each new edge must yield an increase in utility. The HL condition closely tracks the LL condition, because both conditions result from dynamics under low tie costs. 

\begin{figure*}
\includegraphics[width=0.8\textwidth]{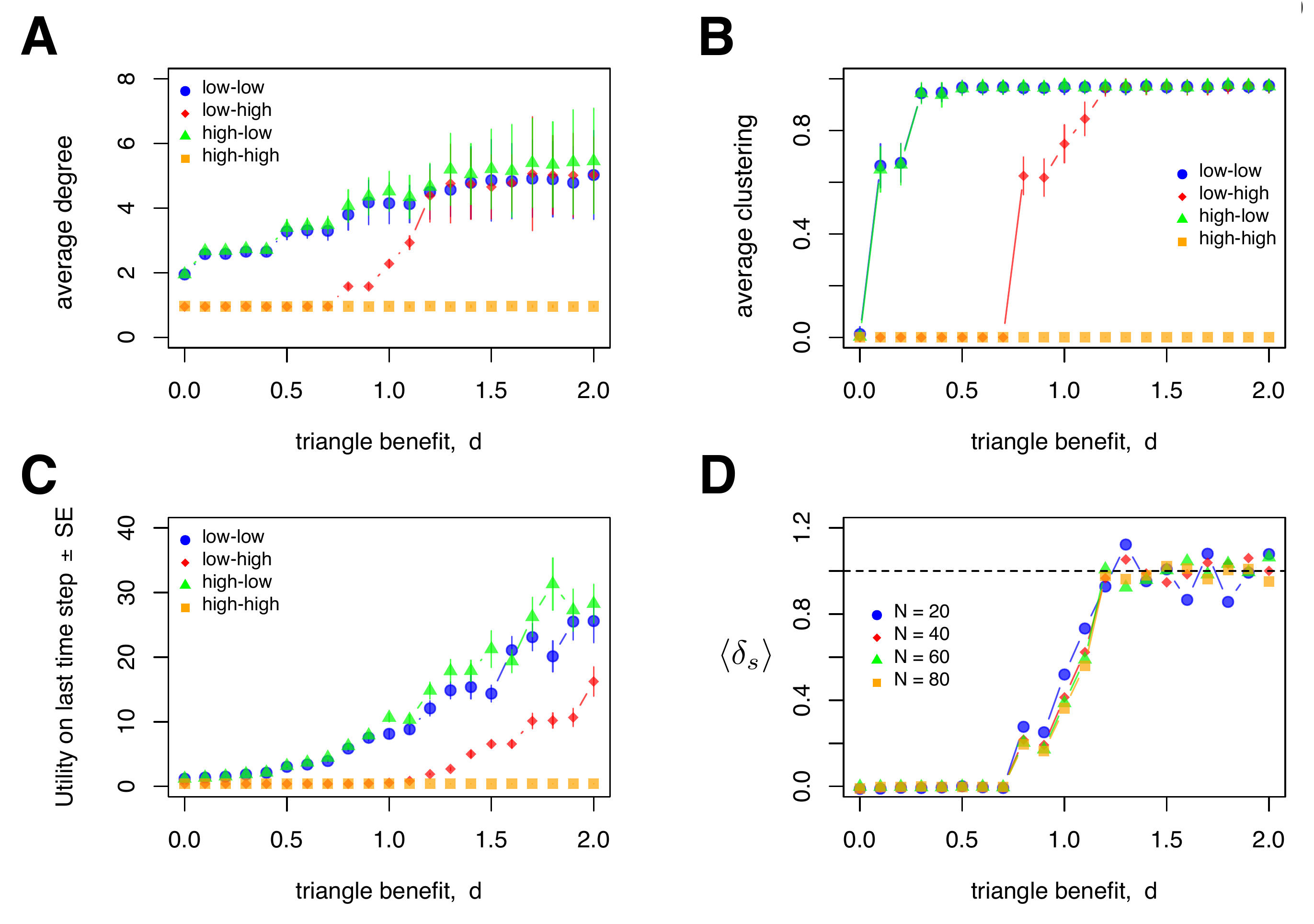}
\caption{\label{fig:fig2} Isolated effects: triangle benefits only ($e = 0$).  (A--C) Average results for each of four shock conditions on a 40-node network. (A) Average node degree $\pm$ SD, (B) Average node clustering $\pm$ SD, (C) Average node utility at equilibrium $\pm$ SE, (D) Average resilience for LH condition, showing insensitivity to network size.}
\end{figure*}

\begin{figure*}
\includegraphics[width=0.9\textwidth]{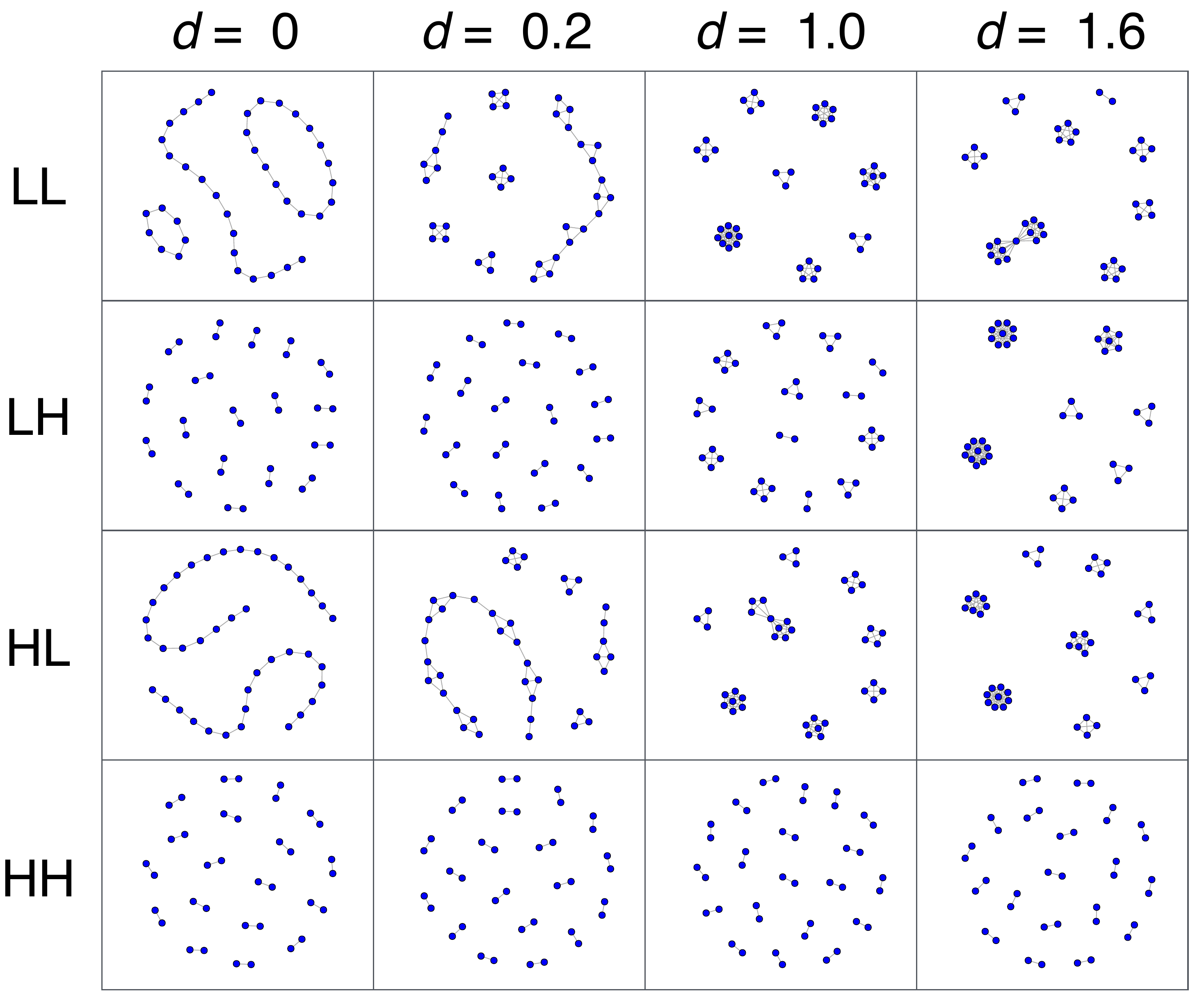}
\caption{\label{fig:fig3} Isolated effects: triangle benefits only ($e = 0$).  Representative single-layer networks that emerge as a result of varying incentives for closed triangles, $d$. Unconnected nodes do occasionally occur but are not represented in these plots.}
\end{figure*}

For values of $d$ below a critical threshold, the LH condition tracks the HH condition. This is because the shock in which tie costs increase causes agents to drop ties, and triangles cannot be maintained. Past the critical threshold ($d = 0.8$ in our runs), some amount of resilience occurs. Some nodes are dropped, but the network is denser than networks that began with high tie costs. This first threshold is when the benefit of a closed triangle can offset the higher tie cost, so that a node in a closed triangle need not drop any ties. 
Past a second threshold ($d = 1.2$ in our runs), when the benefits to closed triangles are high enough, networks in the LH condition are indistinguishable from networks in the LL condition. See below for a derivation of these thresholds. 
Examining the average clustering of the network mirrors this finding (Fig. \ref{fig:fig2}B). When triangles are incentivized and tie costs permit their closure, clustering maximizes fairly rapidly. Fig. \ref{fig:fig3} illustrates the types of network structures that emerge under each shock condition and varying benefits to closed triangles. Examining the average node utility at equilibrium, we also observe that resilience allows post-shock LH nodes to maintain higher utility even after costs increase than they would if costs had always been high.

\subsection{Isolated effects: Spillover benefits only}

\begin{figure*}
\includegraphics[width=0.8\textwidth]{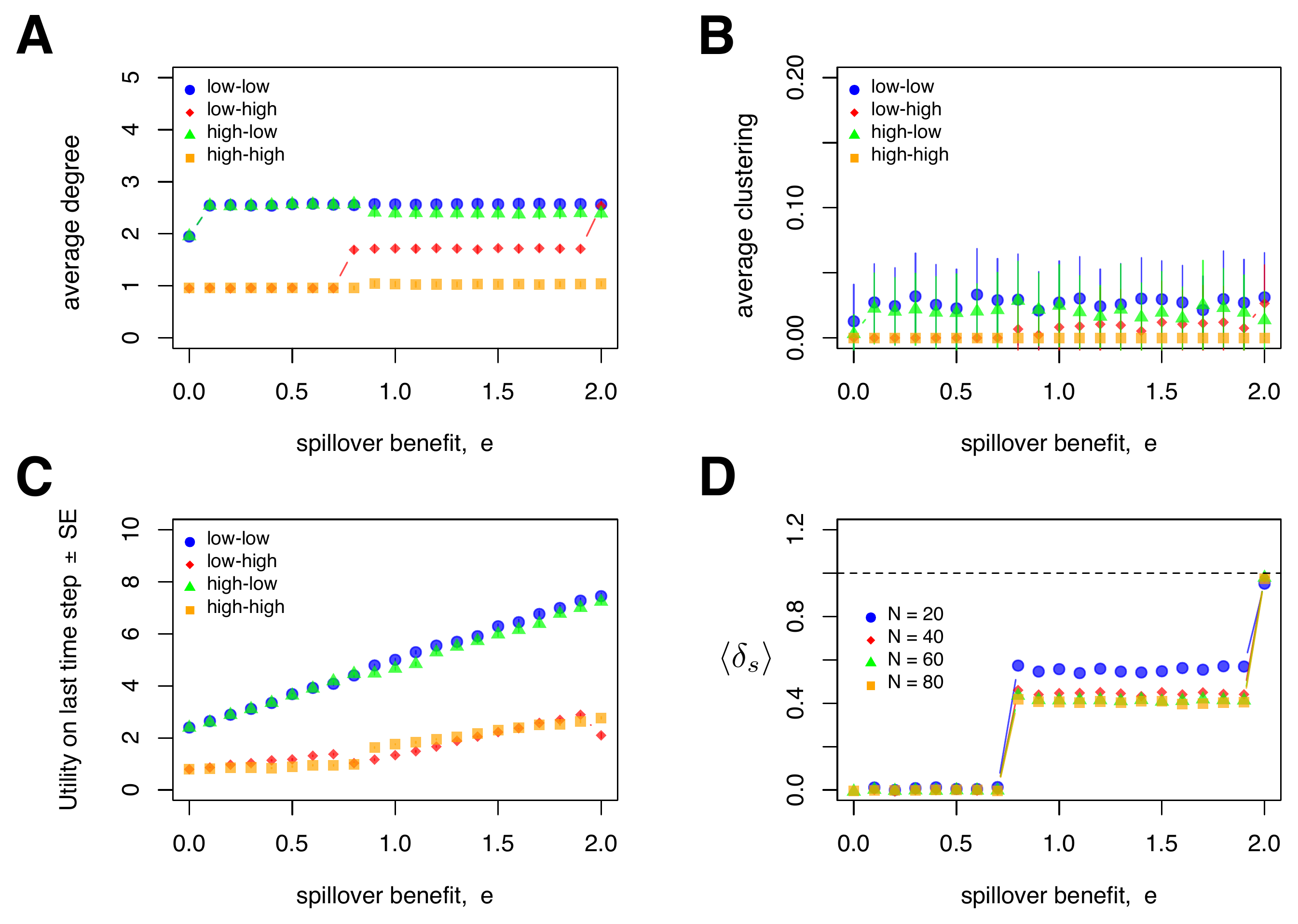}
\caption{\label{fig:fig4} Isolated effects: spillover benefits only ($d = 0$).  (A--C) Average results for each of four shock conditions on a 40-node network. (A) Average node degree $pm$ SD, (B) Average node clustering $\pm$ SD, (C) Average node utility at equilibrium $\pm$ SE, (D) Average resilience for LH condition, showing insensitivity to network size. All but (C) are from Layer 1 only.}
\end{figure*}

We next consider a two-layer multiplex (and will do so for all subsequently presented results). Like incentives for closed triangles in a single-layer network, incentives for spillover ties provide a minimal model of structural entrenchment in a multiplex network. In general, we see a similar pattern of resilience for spillover as we did for triangles (Fig. \ref{fig:fig4}). Unlike with triangles, however, the average degree under low tie costs does not continue to increase with the benefit to spillover ties (Fig. \ref{fig:fig4}A). Rather, it plateaus. This is because spillover ties do not scaffold the creation of additional spillover ties, as closing triangles does. In other words, the existence of a spillover tie does not provide new opportunities for additional spillover ties. The critical threshold for some resilience in the LH condition is the same for spillover as for triangles, which is unsurprising when the benefit of a spillover tie can prevent a node from needing to drop a tie due to increased costs. Unlike with triangles, the network is not fully resilient until a much higher spillover benefit has been reached as compared with the triangle case ($e = 2$ in our runs). This is because each tie can only confer one unit of spillover benefit, whereas a single tie can be part of many triangles. See below for derivation of critical thresholds. 

Past the first critical threshold, the average degree of the HL condition is slightly lower than for the LL condition. This is because all ties will be spillover ties under high costs and large $e$. This ends up making it more difficult for some nodes to find partners who would accept their offer to form a tie. The reason is that fewer would-be partners stand to increase their utility from adding a tie. 
Interestingly, the average utility received by a node at equilibrium in the LH condition is not any higher than that of a node in the HH condition. 
That is, nodes who end up in a high-cost environment experience no benefits nor costs, in the short run, on the basis of whether tie costs were initially high or low. This contrasts the case where we examine variations in triangle benefits. The difference is due to the fact that the benefit of additional spillover ties is compensated by the higher costs of more ties. 

\begin{figure*}
\includegraphics[width=0.9\textwidth]{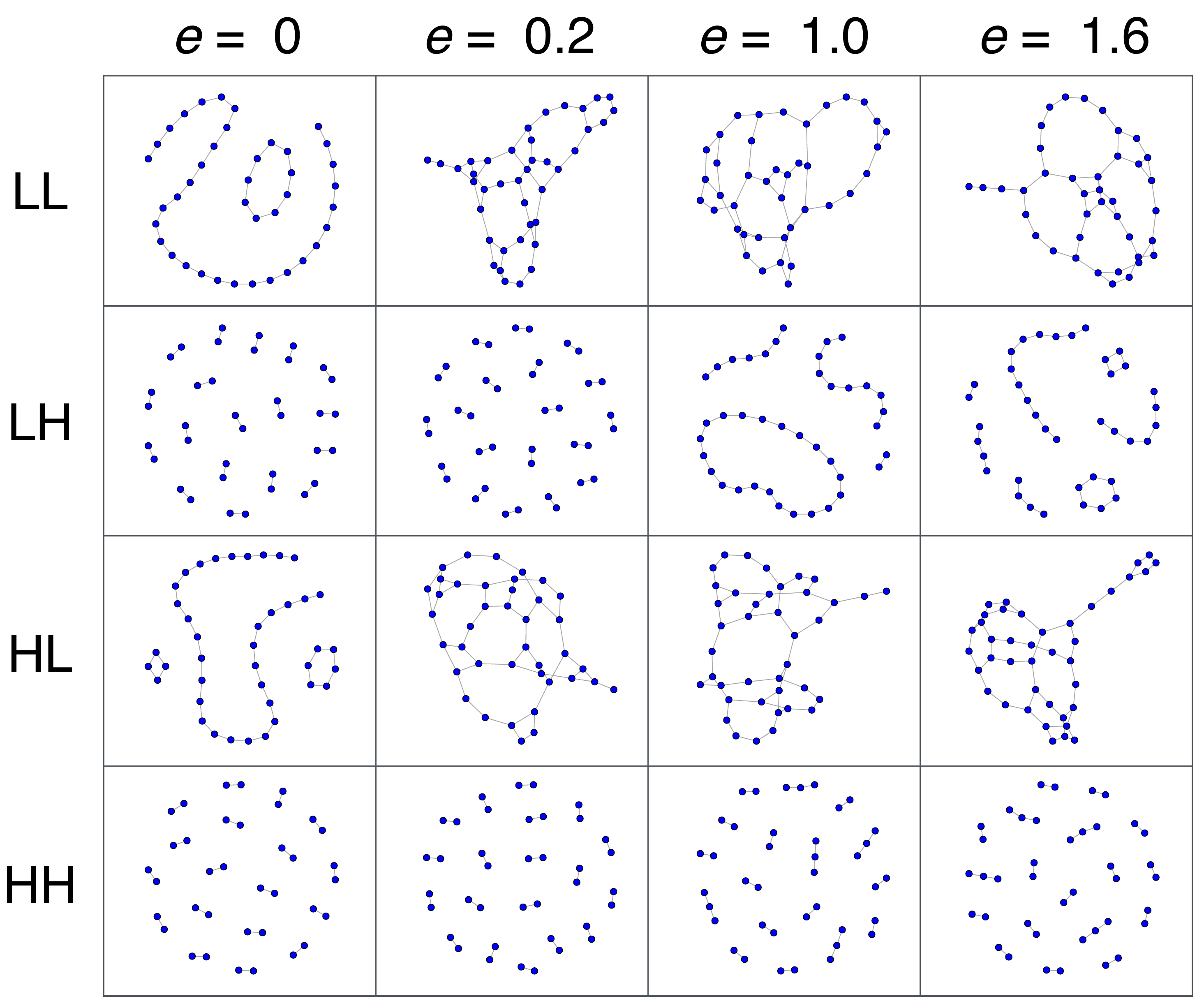}
\caption{\label{fig:fig5} Isolated effects: spillover benefits only ($d = 0$).  Representative networks (Layer 1 only) that emerge as a result of varying incentives for spillover ties, $e$. Unconnected nodes do occasionally occur but are not represented in these plots. }
\end{figure*}

The network structures that emerge from spillover incentives are quite different from those that emerge from triangle incentives (Fig. \ref{fig:fig5}). Under low tie costs, incentives for triangles created several tightly clustered but completely discrete communities. Incentives for spillover, on the other hand, tends to create fully connected graphs that exhibit low levels of triadic closure (Fig. \ref{fig:fig4}B). 
This structure is not fully recovered in the LH shock condition. Rather, an intermediate structure emerges composed of several isolated chains or circles. 

\section{Explanation of Transition Points}

\subsection{When resilience begins.}
Here we derive conditions for when, on average, $k_{LH} > k_{HH}$. We consider only the isolated effects conditions for clarity. In addition, we focus on a minimal type of resilience observed in our simulations: when two or more edges are not possible under high tie costs but are present following a shock to high costs from initially low tie costs. Different types of resilience for different degree thresholds are also possible, as shown in later sections of this Appendix.

\begin{figure*}
\includegraphics[width=0.7\textwidth]{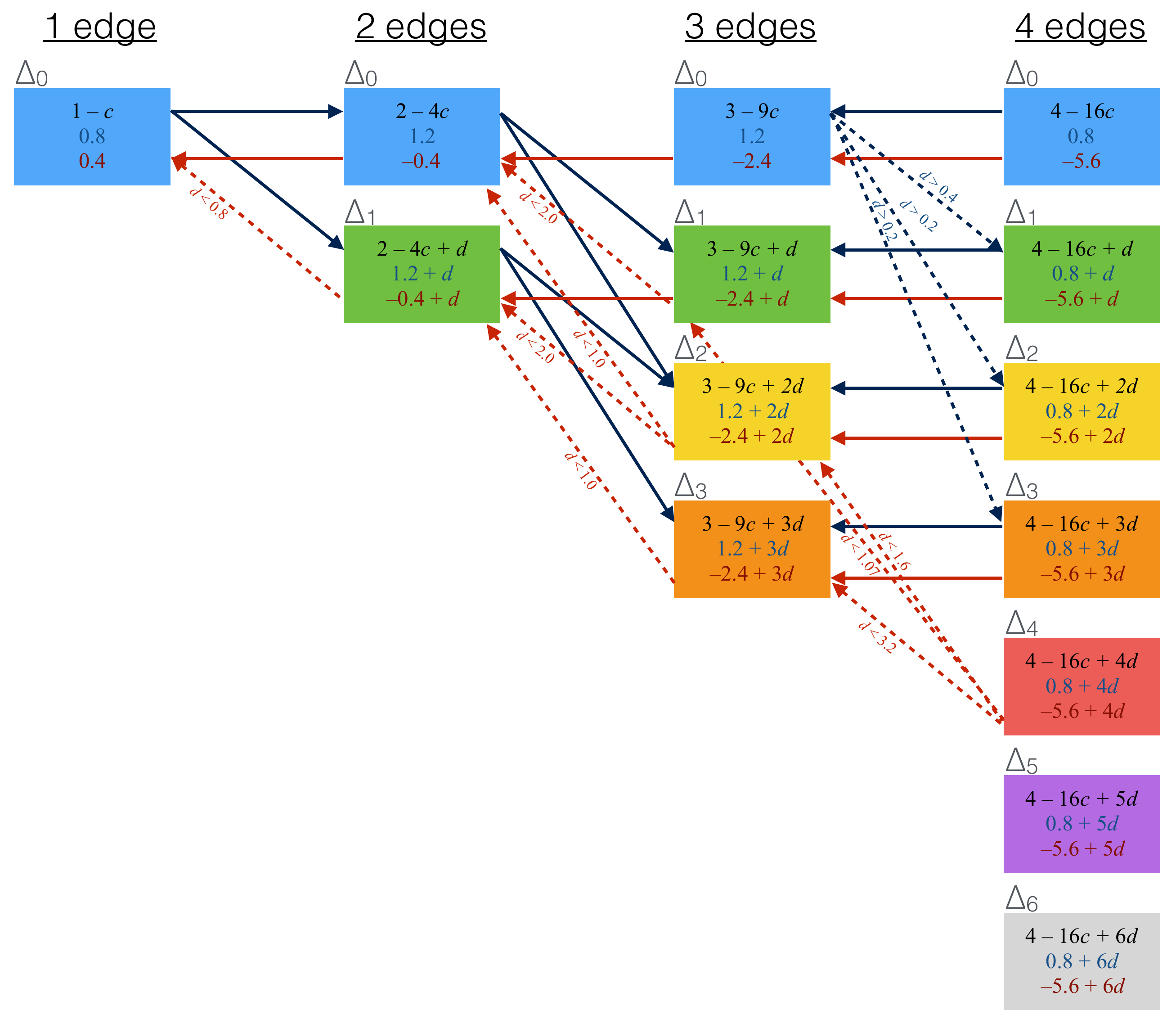}
\caption{\label{fig:figS1} Individual utilities for triangle benefits only ($e = 0$), for different number of ties and triangles ($\Delta$), and utility-increasing state transitions under LH shock conditions. Precise values are given for low tie costs, $c_{low}$, and high tie costs, $c_{high}$, in blue and red, respectively. Blue arrows indicate when adding (or dropping) an edge would result in a utility increase under low tie costs. Red arrows indicate where dropping an edge would be incentivized under a post-shock tie-cost increase. Solid lines indicate a move that is always favored (sometimes only when $d > 0$), dashed lines indicate moves dependent on the triangle benefit, $d$. For the transition between 3 and 4 ties, only a subset of transition lines are shown for clarity. The remaining transitions can be easily calculated with the values shown.}
\end{figure*}

First, let us condition only triangle benefits ($e = 0$) under a single-layer network. Figure \ref{fig:figS1} indicates the threshold parameter values for additional ties. We see that under low tie costs, there is always the incentive to have at least two social ties, while under high costs, only one tie is incentivized. Under low tie costs, the utility for two and three ties is identical, and so adding a third tie only occurs if $d > 0$, which is what we observed (main text, Fig. 2). When tie costs increase from low to high, we see that the triangle benefits must be quite high to maintain three ties, unless the individual node already has three triangles. Thus, there is often a reduction from four or three ties to two. However, two ties can be stable as long as $d \geq 0.8$ and the agent is in a closed triangle, because only when it is below this threshold is there a strict increase in agent utility from dropping an edge. More generally, this minimal level of resilience between two and one network ties will be seen when the following two conditions are met: (1) a second edge will never be added {\em de novo} under high tie costs but will always be favored under low tie costs, and (2) the benefit to triangles ensures that, if a closed triangle exists, dropping an edge, and hence losing the triangle, will not be favored under either ties cost level. 

Condition 1 is met when 
\[
1 - c_{low} < 2 - 4c_{low},
\]
or when $c_{low} < 1/3$, and, correspondingly, $c_{high} \geq 1/3$.  

Condition 2 is met when 
\[
d \geq 3c_{high} - 1.
\]
Under the value we used, $c_{high}$, the threshold value of $d$ is 0.8, which is exactly what we observed in our simulations. 

\begin{figure*}
\includegraphics[width=0.95\textwidth]{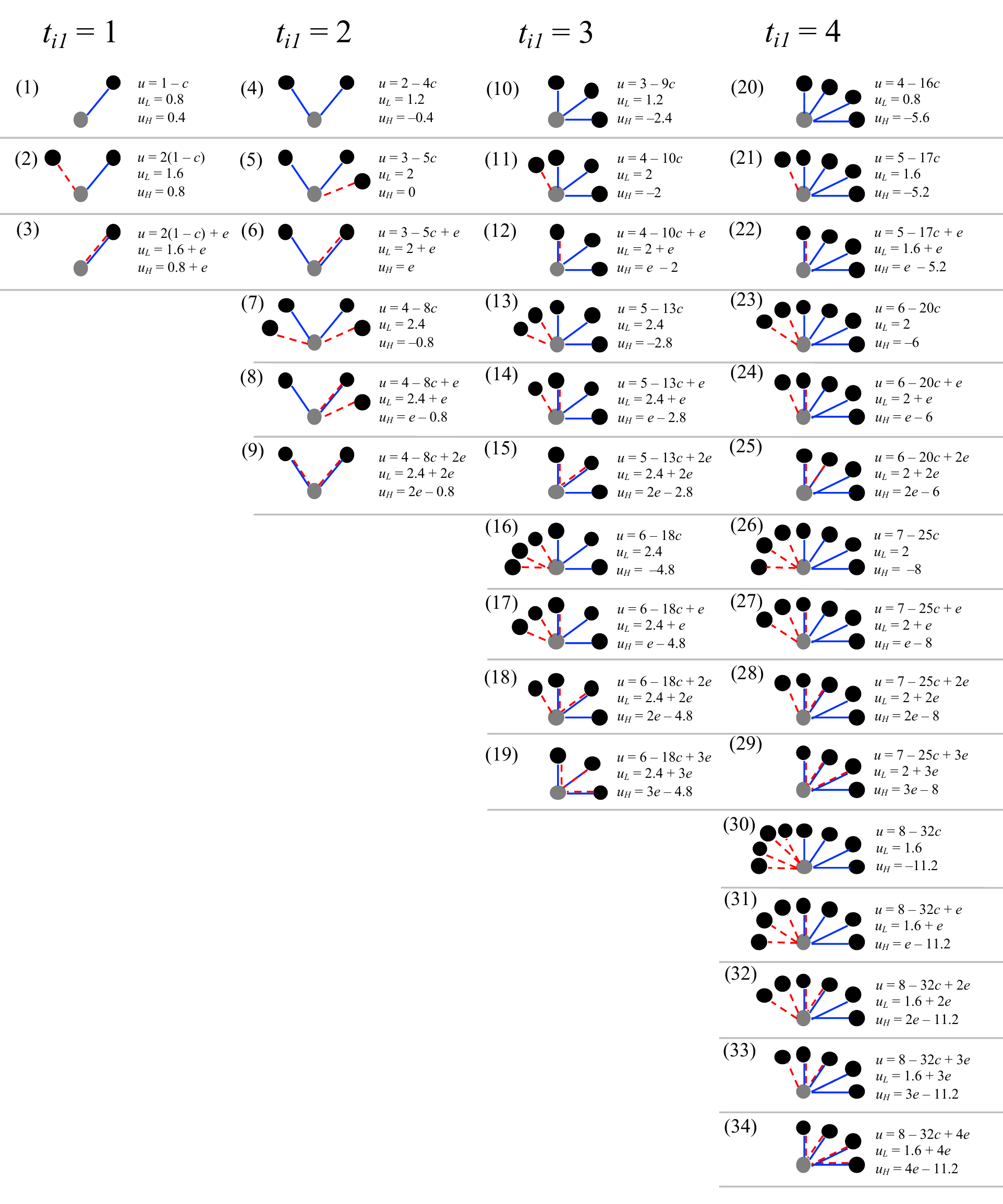}
\caption{\label{fig:figS2} All 34 possible ego networks (centered on the grey node) and corresponding utilities for spillover benefits only ($d = 0$). Ties in layer 1 are indicated by solid blue lines, ties in layer 2 are indicated by dashed red lines. Utility values are given for low tie costs ($u_L$) and high tie costs ($u_H$), based on the values for  $c_{low}$ and $c_{high}$ used in the main text. A transition diagram for these networks under each cost and shock condition (LH, HL), similar to that shown in Figure \ref{fig:figS1}, can be derived from these states and corresponding utilities, though it will be considerably more complicated.} 
\end{figure*}

The logic of this analysis is easily extended to the case of spillover benefits only ($d = 0$, $e > 0$), although there are are a greater number of relevant ego networks, making the transition diagram quite complicated. See Figure \ref{fig:figS2}. The presence of a spillover benefit for either of a node's ties allows a second tie to be maintained after a shock from low to high tie costs. This logic also explains the diagonal threshold line seen in Figure 4B (main text) in the LH shock condition. The resiliency effects of triangle and spillover benefits are additive, such that if an agent possesses {\em both} a spillover edge and a closed triangle, it is resilient to shocks as long as the total $d + e$ is greater than the threshold, which in this case is 0.8. 

\subsection{When resilience is perfect.}
When does $k_{LH} = k_{LL}$? In our simulations, we observe that, when triangle or spillover benefits are sufficiently large, resilience is perfect, and the average degree of the network does not diminish when a network formed under low tie costs experiences a sudden increase to tie costs. What this means is that the incentives are such that a stable state reached under low tie costs will not become unstable when tie costs are suddenly made high. 

This is most easily illustrated by considering the case of spillover benefits only ($d = 0$; see Figure \ref{fig:figS2}). Under low tie costs, nodes will often reach degree 4 (Note that this is the average degree for Layer 1 only. When the layers have the same incentives, network statistics are the same for both layers.). However, such a high degree is unstable unless all edges are spillover edges. And because it is not always possible to increase the degree of a node and increase the number of spillover edges simultaneously, degree 3 (and even degree 2) is the common and stable network state for low tie costs (see main text Figures 4 and 5). Although degree 3 is stable under low tie costs, it is unstable after post-shock high tie costs unless $e \geq 2.0$, which is the threshold point for perfect resilience. If $e$ is less than this, a node's degree will decrease to $k= 2$ if $e \geq 0.8$ (partial resilience), and $k = 1$ otherwise (no resilience). This argument can be extended for all shock-related network dynamics for any incentive parameter values.

\subsection{When complete spillover occurs in HH condition.}

\begin{figure*}
\includegraphics[width=0.4\textwidth]{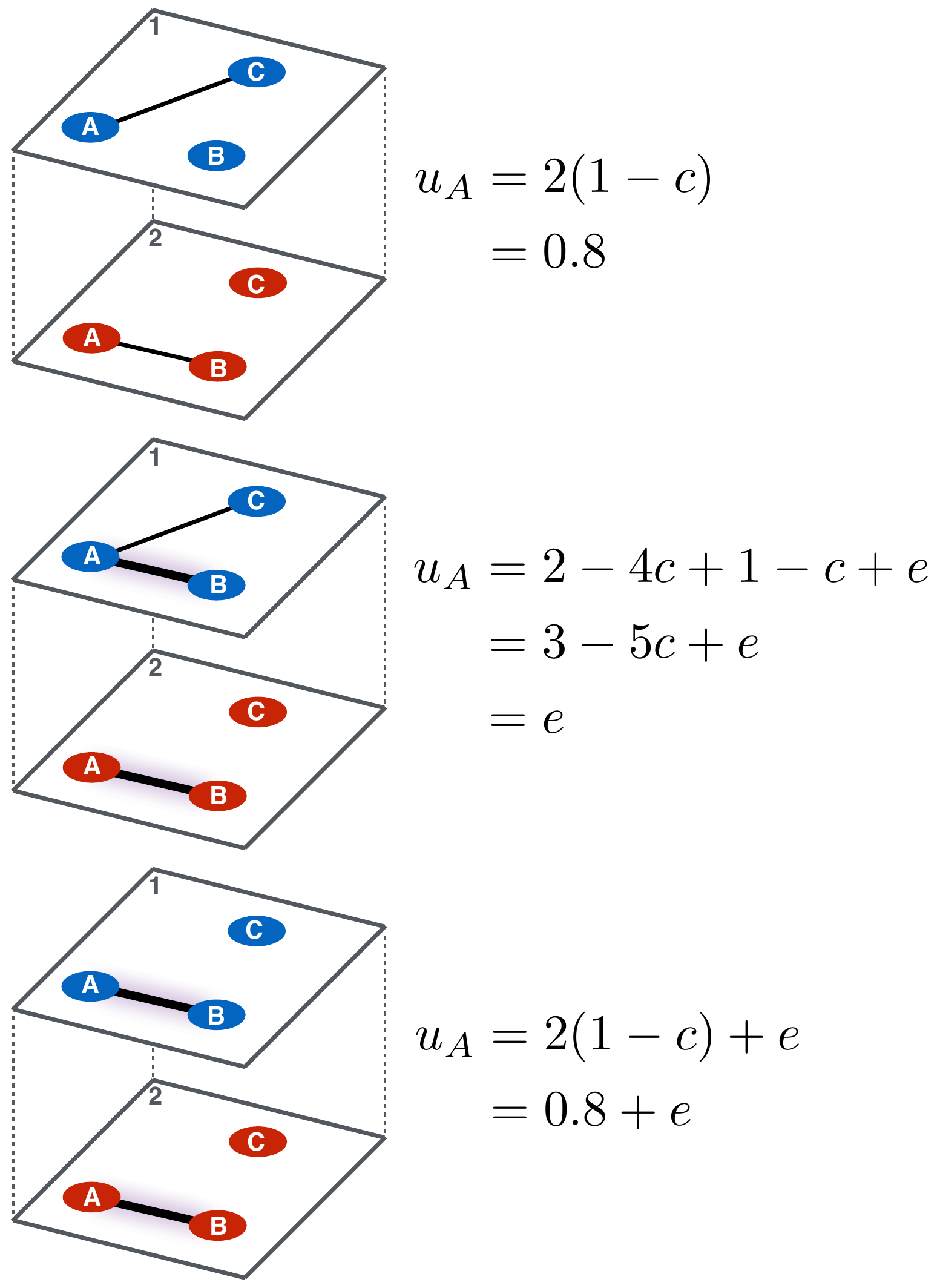}
\caption{\label{fig:figS3} The dynamics of spillover under high tie costs. Consider an agent $A$ who has a tie with agent $C$ in layer 1 and a tie with agent $B$ in layer 2 (top row).The agent can add a tie in one of the two layers to complete a spillover edge; in this case with agent $B$ in layer 1 (middle row). Such a move is favored if the utility gained from the new spillover tie is great enough to compensate the added costs of the second tie in layer 1; in our case, this occurs when $e > 0.8$. Once this new tie is formed, it is then always beneficial to agent $A$ to drop its previously held tie with agent $C$ in layer 1 (bottom row). Thus, under high tie costs, we observe a threshold value of $e$ above which the average degree does not change (it remains $k = 1$), but for which the proportion of spillover edges increases to unity. Below this threshold, there are still more spillover ties than expected from random assortment, due to limited incentives to form spillover ties. Utilities calculated assume our simulation value of $c_{high} = 0.6.$}
\end{figure*}

Under constant high tie costs (HH condition), there is a threshold value of the spillover benefit, $e$, above which all ties are spillover ties (see Figure 8 in the main text). For our simulations, this value is $e > 0.8$. Such a state occurs when the cost of adding a new tie that completes a spillover edge is favored, but subsequently dropping any non-spillover ties is also favored. This is illustrated in Figure \ref{fig:figS3}.

\section{Probability of spillover pairs from random pairing}
Earlier in this Appendix, we calculated the threshold transition parameter for when all ties will be spillover ties. Before that, our simulations indicate that a smaller number ties are spillover ties, except with $e = 0$, for which spillover ties are rare. In such a case, spillover ties are rare due to the fact that they will only occur by chance, and the number occurring may be less than expected in a purely random model due to high numbers of isolated clusters that form when only triangle benefits are present. In other cases, spillover ties follow a pattern in which they are weakly incentivized, and therefore the proportion corresponds to numbers higher than should be expected by chance. What is this number? 

We can calculate this for the special case in which each node has degree of 1, which occurs under high tie costs. Under random pairing, each node chooses the name node as its neighbor in each layer with probability $1/(N-1)$, and this is equal to the expected proportion of edges that will co-occur in both layers, i.e., the proportion of spillover edges. For a 40-node network, as was used in most simulations presented in the main text, this approximately equal to $0.026$.

\section{Explanation of Fig. 4B in the main text}
\begin{figure*}
\includegraphics[width=0.95\textwidth]{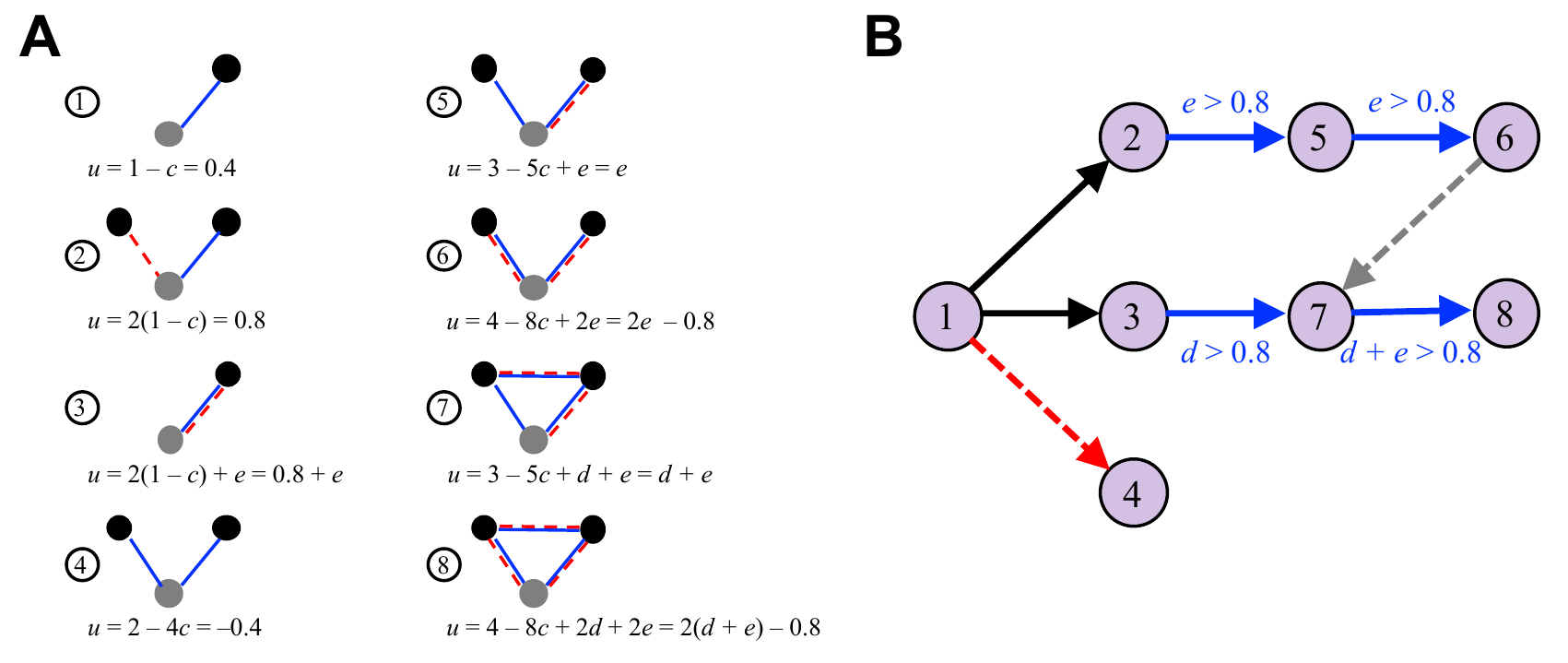}
\caption{\label{fig:figS12} (A) Eight distinct ego networks (for the grey node) under high tie costs ($c_{high} = 0.6$), with utilities indicated. Edges can exist both in layer 1 (blue) and layer 2 (red) of the multiplex.  (B) Transitions between the states indicated in subfigure A. Black arrows indicate transitions that are always favored. Blue arrows indicate transitions that are sometimes favored; the required condition for each transition is shown in blue text. The grey dashed arrow between state 6 and state 7 indicates that the presence of state 6 centered on an adjacent node is required for ego to transition from state 3 to state 7. The red dashed line indicates that this transition is never favored under the indicated cost condition. }
\end{figure*}

Figure 4B, bottom row, in the main text shows a curious result: the resilience of under HL shocks is lower when structural benefits are significantly high---specifically when $d, e > 0.8$. To explain this we note that under such high structural incentives, many nodes will form triangles even under high tie costs. Thus, when tie costs are lowered, there are fewer new connections that will be incentivized. Figure \ref{fig:figS12} provides an illustration of how this can occur. 

\section{Supplemental Simulation Results}

\subsection{Sensitivity to noise}
When network dynamics exhibit resilience, post-shock equilibria are metastable in LH conditions, befitting the path-dependent nature of the equilibria (i.e., the network states cannot be obtained from an empty network). As such, random events---adding or dropping edges at random---will eventually eliminate resilience, causing the system to settle into a state resembling those obtained under high initial tie costs. The key word here is {\em eventually.} To investigate the time scale of these dynamics, we ran simulations in which adds and drops occurred with probability $\nu$ (see details in main text).  We found that after shocks from low to high tie costs, the system moved from the metastable (LH) higher-degree state to the stable (HH) low-degree state at a timescale that was approximately $t \sim 1/\nu$. This was confirmed for $\nu \in [10^{-4}, 10^{-1}]$. Our results therefore hold as long as most events are strictly utility-increasing, relative to the characteristic timescale of dynamics.  

\begin{figure*}
\includegraphics[width=0.4\textwidth]{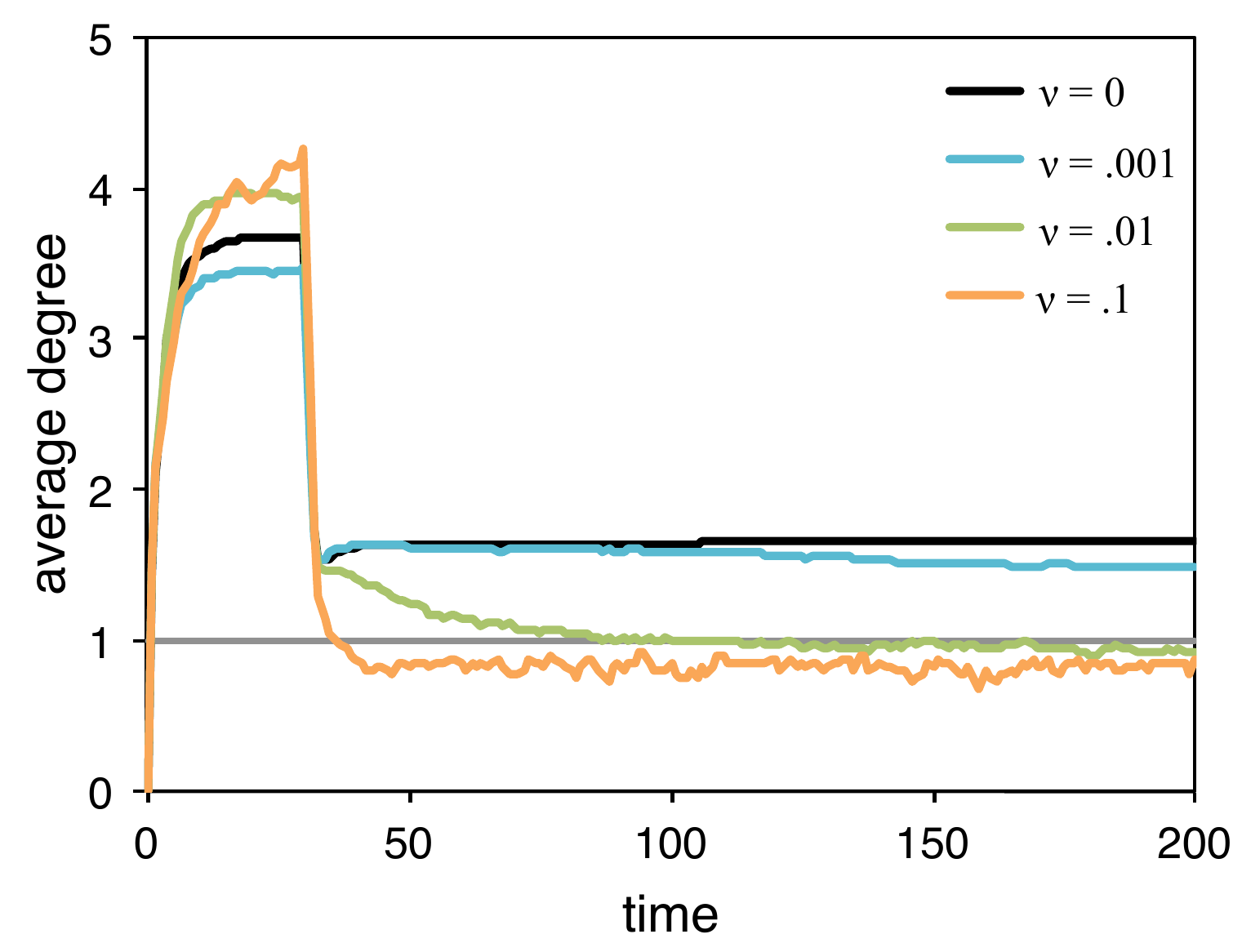}
\caption{\label{fig:figS4} Sensitivity to noise. Temporal dynamics of representative simulation runs. Under initially low tie costs, average degree increases to a dynamic equilibrium of about 3 or 4.  A shock to high tie costs occurs at $t = 40$, after which we see a decrease in average degree. In the absence of noise, this stabilizes to an average degree of just under 2. The more noise is present, the more quickly the system goes from the metastable $LH$ condition (black line) to the stable $HH$ condition (grey line).}
\end{figure*}

\subsection{Sensitivity to population size}
\begin{figure*}
\includegraphics[width=0.9\textwidth]{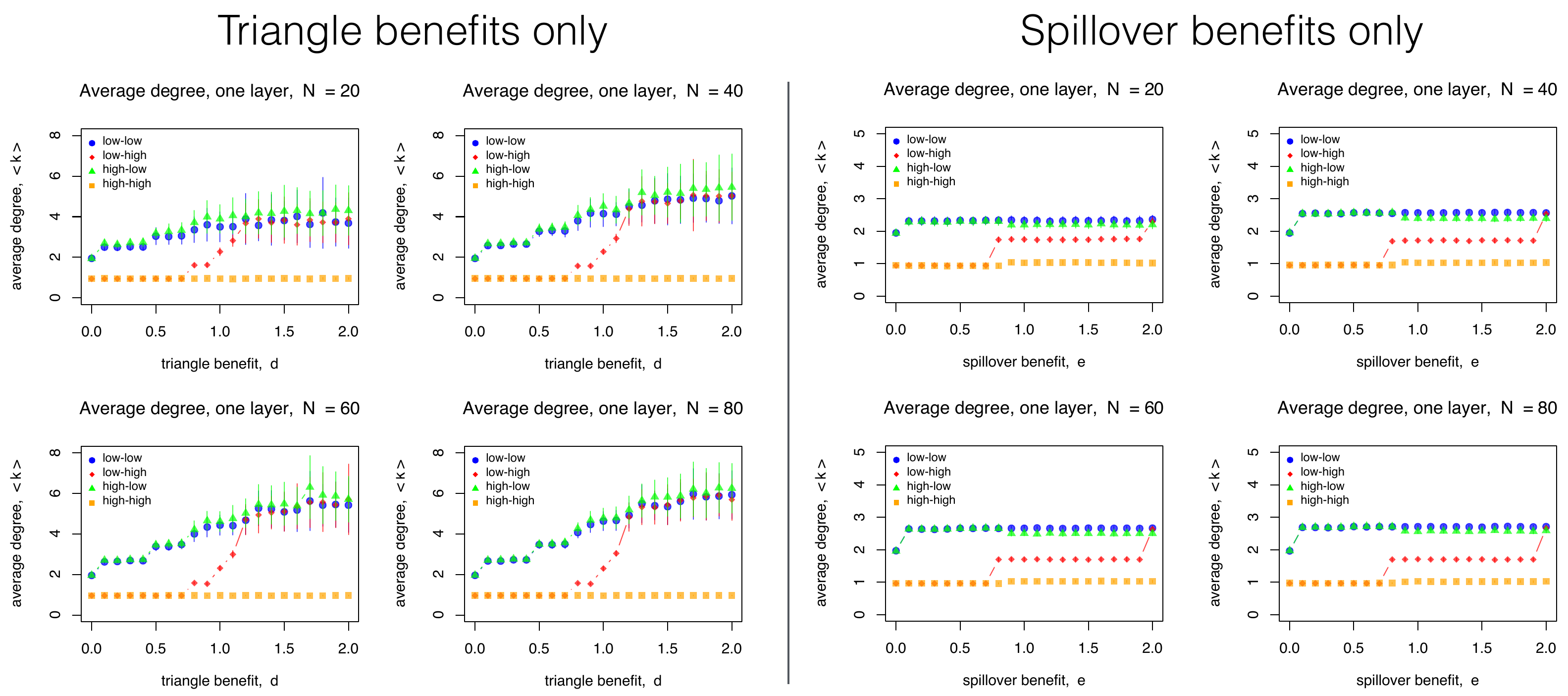}
\caption{\label{fig:figS5} Sensitivity to network size: average degree.}
\end{figure*}

\begin{figure*}
\includegraphics[width=0.9\textwidth]{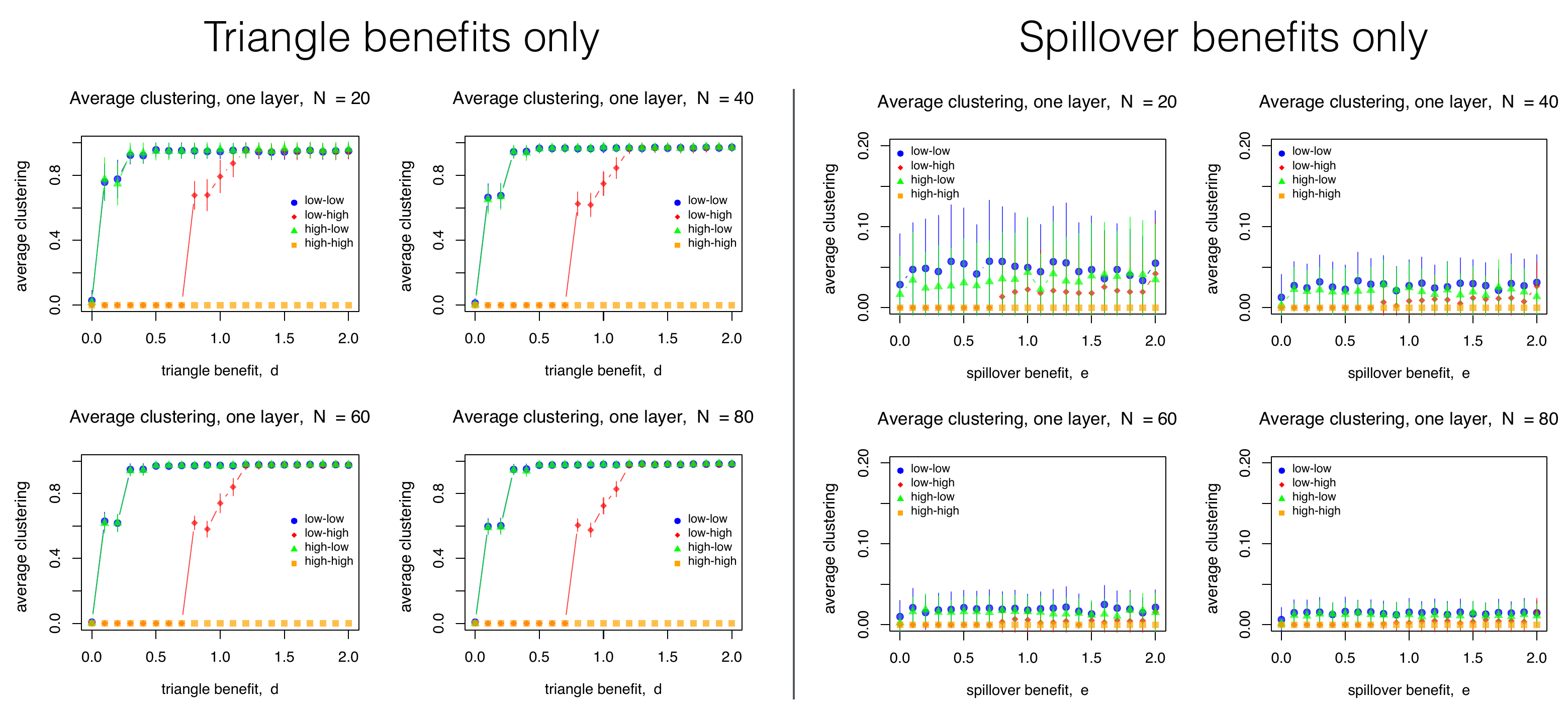}
\caption{\label{fig:figS6} Sensitivity to network size: average clustering.}
\end{figure*}

Our results were very robust to changes in population size. This is largely because the numerical values of individual incentives operated on the tie capacity of nodes, independent of the size of the network. Larger network created few shortages for social ties, and the so the average degree of agents tended to be slightly higher in larger networks than in smaller networks, but this affect was minimal (Figure \ref{fig:figS5}). Average clustering was similarly robust (Figure \ref{fig:figS6}). For triangle benefits only, clustering was slightly higher in very small networks, due to more triangles forming through random chance as a result of the small population. The same was true for the spillover benefits only case.  In this case, incentives tended to push the network {\em away} from clustering. When network size was very small, some additional clustering happened as a result of change connections. This effect disappeared for larger networks. 

\subsection{Sensitivity to tie costs}

\begin{figure*}
\includegraphics[width=0.6\textwidth]{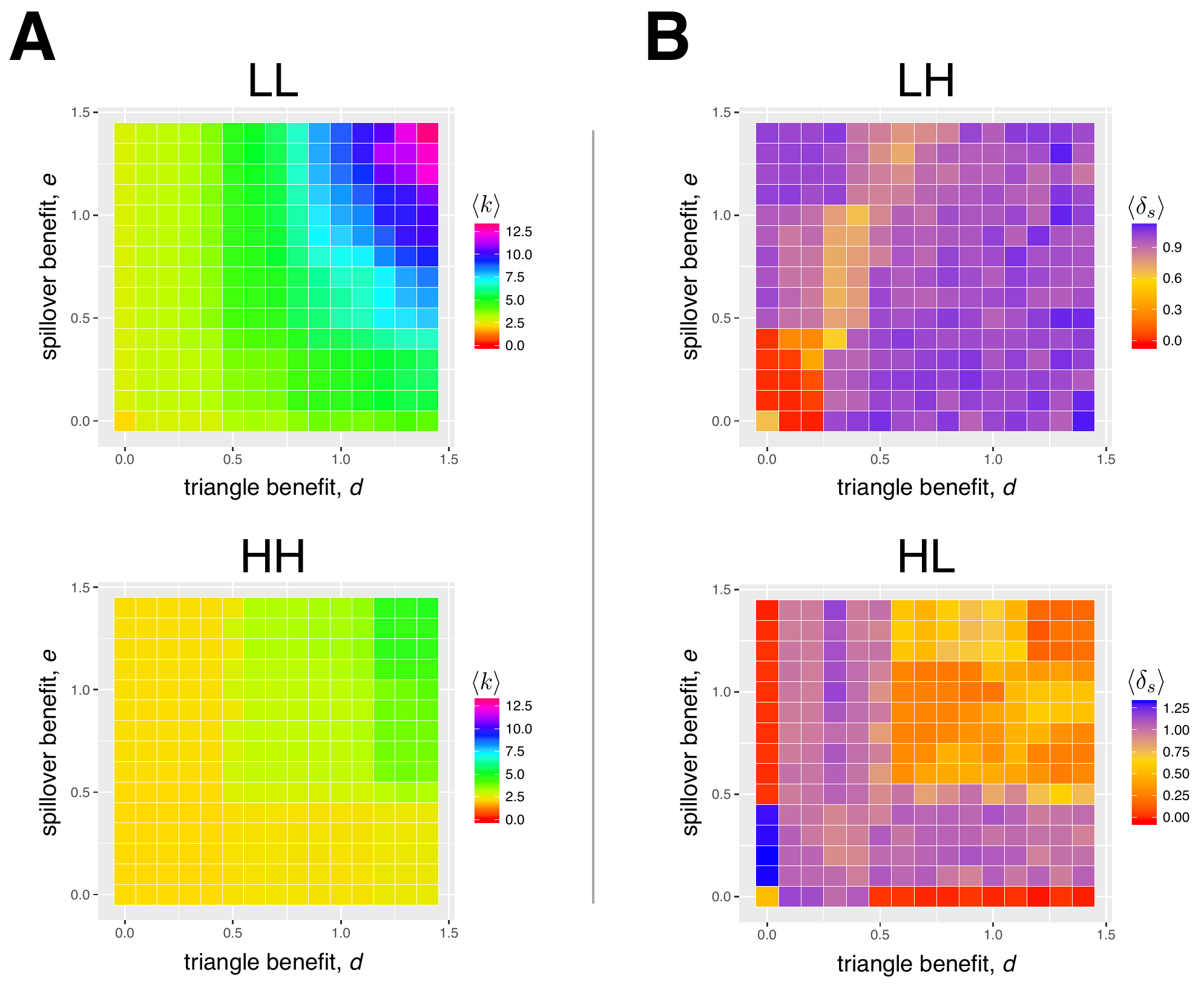}
\caption{\label{fig:figS7} Average degree for all four shock conditions when $c_{high} = 0.3$ and $c_{low} = 0.2$.}
\end{figure*}

The results shown in the main text used parameters chosen for maximal clarity. For example, when tie costs were always high (HH condition), the equilibrium degree was exactly one. However, the broader principle of our results---namely, resilience from structural entrenchment---should hold for a wide range of parameters. To demonstrate this, we rand simulations for which the ``high" cost of social ties was sufficiently low to generate higher degree networks, and thus the possibility of triangles. Figure \ref{fig:figS7} illustrates that, although the resilience effects are less stark, there are similar patterns of resilience as seen with more extreme tie cost values.

\newpage

\end{document}